\documentclass[twocolumn,tighten,times]{aastex62}
\usepackage{graphicx,xfrac,amsmath, color}
\usepackage{float}
\usepackage[caption = false]{subfig}
\usepackage{rotating}
\usepackage{caption}
\usepackage[T1]{fontenc}
\usepackage{ae,aecompl}

\received{2023 March 31}
\submitjournal{Astronomy \& Geophysics, June 1 2003, Volume 64, Issue 3}

\shorttitle{Betelgeuse}
\shortauthors{Wheeler \& Chatzopoulos}

\begin{document}
\def\la{\mathrel{\hbox{\rlap{\hbox{\lower4pt\hbox{$\sim$}}}\hbox{$<$}}}}
\def\ga{\mathrel{\hbox{\rlap{\hbox{\lower4pt\hbox{$\sim$}}}\hbox{$>$}}}}
\def\mic{$\mu$m}
\def\mch{M$_{Ch}$}
\def\lam{$\lambda$}
\def\kms{km~s$^{-1}$}
\def\cms{cm~s$^{-1}$}
\def\vphot{$v_{\rm phot}$}
\def\ang{\AA}
\def\syn{SYNOW}
\def\dm15{{$\Delta$}$m_{15}$}
\def\rsi{$R$(Si~II)}
\def\v10{$V_{10}$(Si~II)}
\def\wsi{$W_lambda$(Si~II)}
\def\vdot{$\.v$(Si~II)}
\def\W575{$W(5750)$}
\def\W610{$W(6100)$}
\def\6100{the 6100~\AA\ absorption}
\def\tex{$T_{\rm exc}$}
\def\ve{$v_{\rm e}$}
\def\magd{mag~d$^{-1}$}
\def\m{M$_\odot$}
\def\msun{M$_\odot$}
\def\msunyr{M$_\odot$~y$^{-1}$}
\def\Msun{M$_{\rm \odot}$}
\def\mni{$M_{Ni}$}
\def\Lsun{L$_\odot$}
\def\rsun{R$_\odot$}
\def\ni{$^{56}$Ni}
\def\co{$^{56}$Co}
\def\fe{$^{56}$Fe}
\def\ti{$^{44}$Ti}
\def\ho{H\"oflich}
\def\Ha{H$\alpha$\,}
\def\Hb{H$\beta$\,}
\def\Hg{H$\gamma$\,}
\def\etal{et~al.}
\def\arcsec{$^{\prime\prime}$}
\def\ergs{erg~s$^{-1}$}
\def\gcm3{g~cm$^{-3}$}
\def\cm2g{cm$^{2}$~g$^{-1}$}
\def\Po{$P_{\rm o}\ $}
\def\degree{$^{\rm o}$}
\def\CaII7291{[Ca~{\sc II}] $\lambda\lambda$7291,7323}
\def\HeIA{He~{\sc I} $\lambda$5876}
\def\HeIB{He~{\sc I} $\lambda$6678}

\def\OI6300{[O~{\sc I}] $\lambda\lambda$6300,6364}

\def\OIA{[O~{\sc I}] $\lambda$ 5577}

\def\OIB{[O~{\sc I}] $\lambda$ 7774}

\def\NaID{Na~{\sc I}~D}
\def\MgI{Mg~{\sc I}] $\lambda4571$}
\def\qisp{Q$_{ISP}$}
\def\uisp{U$_{ISP}$}
\def\fluxu{${\rm ergs\, cm^{-2} \AA^{-1} s^{-1}}$}
\def\uisp{U$_{ISP}$}
\def\fluxu{${\rm ergs\, cm^{-2} \AA^{-1} s^{-1}}$}
\newcommand\gr{$\gamma$--ray}
\newcommand\grs{$\gamma$--rays}
\newcommand\grb{gamma--ray burst}
\newcommand\grbs{gamma--ray bursts}
\def \lta {\mathrel{\vcenter
     {\hbox{$<$}\nointerlineskip\hbox{$\sim$}}}}
\def \gta {\mathrel{\vcenter
     {\hbox{$>$}\nointerlineskip\hbox{$\sim$}}}}
\def\apj{ApJ}
\def\apjl{ApJL}
\def\apjs{ApJS}
\def\aj{AJ}
\def\nat{Nature}
\def\mnras{MNRAS}
\def\aap{A\&A}
\def\pasp{PASP}
\def\apss{ApSpSci}
\def\iaucirc{IAU Circ.}
\def\actaa{Acta Astron}
\def\nar{New Astron Rev}
\def\araa{ARAA}
\def\pasj{PASJ}

\newcommand{\Rsun}{\ensuremath{R_{\odot}}}
\newcommand{\Mdot}{\ensuremath{\dot{M}}}
\newcommand{\MCh}{\ensuremath{M_{\rm Ch}}}
\newcommand{\MWD}{\ensuremath{M_{\rm WD}}}
\newcommand{\RWD}{\ensuremath{R_{\rm WD}}}
\newcommand{\Mmax}{\ensuremath{M_{max}}}
\newcommand{\viscKH}{\ensuremath{\nu_{\rm KH}}}
\newcommand{\viscBC}{\ensuremath{\nu_{\rm BC}}}
\newcommand{\viscZ}{\ensuremath{\nu_{\rm Z}}}
\newcommand{\verthat}{\ensuremath{{\bf \hat{r}}}}
\newcommand{\horihat}{\ensuremath{{\bf \hat{\theta}}}}
\newcommand{\zeehat}{\ensuremath{{\bf \hat{z}}}}
\newcommand{\azihat}{\ensuremath{{\bf \hat{\phi}}}}
\newcommand{\wtot}{\ensuremath{\langle w_{\rm tot} \rangle}}

\def \etc{{\sl etc.}}
\def \ie{{\sl i.e.}}
\def \eg{{\sl e.g.}}
\def \viz{{\sl viz.}}
\def \etal{{\sl et al.}}

\newcommand{\insertfig}[4]{
  \begin{figure}[!htbp]
   \begin{center}
      \includegraphics[#1]{#2}
   \end{center}
   \vspace{-0.00cm}
   \caption{#3}
   \label{#4}
  \end{figure}
}

\newcommand{\insertdoblfig}[6]{
  \begin{figure}[!htbp]
   \begin{center}
     \includegraphics[#1]{#2}
     \includegraphics[#3]{#4}
   \end{center}
   \vspace{-0.00cm}
   \caption{#5}
   \label{#6}
  \end{figure}
}

\title{Betelgeuse: a Review}

\author{J.\ Craig Wheeler}
\affiliation{Department of Astronomy, University of Texas at Austin, Austin, Texas, USA}

\author{Emmanouil Chatzopoulos}
\affiliation{Department of Physics \& Astronomy, Louisiana State University, Baton Rouge, 70803, Louisiana, USA}
\affiliation{Institute of Astrophysics, Foundation for Research and Technology-Hellas (FORTH), Heraklion, 70013, Greece}

\correspondingauthor{J. Craig Wheeler}
\email{wheel@astro.as.utexas.edu}

\begin{abstract}



Betelgeuse has fascinated people since they first looked at the sky. Here we present a contemporary summary of the observations and theory that lead to current understanding of Betelgeuse as a massive red supergiant doomed to eventual collapse and explosion, probably $\sim$ 100,000 years from now. Although it lies only $\sim 200$ parsecs from Earth, and hence can be spatially resolved with appropriate instrumentation, uncertainties in its distance remain a critical impediment to deeper understanding.

The surface of Betelgeuse is rent with a complex structure as deep convective eddies arise to the surface affecting the photosphere, chromosphere, mass loss, the formation of dust and molecules, and the surface magnetic field structure. The global effective temperature has some irreducible uncertainty because of associated temperature variations in the atmosphere. The surface gravity is not precisely known, leading to further uncertainties in the current mass. Determination of the equatorial rotation velocity is critical since some current estimates indicate that Betelgeuse is rotating anomalously rapidly, near rotational breakup, a property that cannot be explained by basic single-star evolutionary models. Betelgeuse is also moving through space at high, though not unprecedented, velocity that indicates that it received a boost, perhaps through collective interaction with other stars in its birth cluster, though disruption of an original binary system has been suggested. A bow shock and other structure in the direction of the motion of Betelgeuse suggests that it has affected the organization of the distant circumstellar and interstellar medium. Betelgeuse varies in brightness on a variety of time scales with $\sim 200$ d, $\sim 400$ d and $\sim 2000$ d being prominent. Models of this variability may be in conflict with historical records suggesting that Betelgeuse was yellow in color, not red, only two millennia ago.

Betelgeuse is also subject to a rich variety of theoretical studies that attempt to understand its observational properties and current evolutionary state. Betelgeuse is statistically probable to have been born in a binary system, and the high space velocity and apparent rotation have been related to binary star evolution. One possibility is that Betelgeuse has been subject to common envelope evolution in which a companion star plunges into the primary and becomes tidally disrupted as it nears the core of the primary. This interaction is complex in three dimensions and not sufficiently well understood. Such merger models have been invoked to account for the apparently anomalous rotation velocity.

Betelgeuse underwent a Great Dimming in 2020 that caught the attention of astronomers and the general public world wide. Explanations have focused on large cool spots on the surface and the expulsion of a cloud of dust that obscured the surface.

We finally sketch the nature of the explosion to come and finish with perspectives for further research.
\\
\\
\end{abstract}



\section{Introduction}
\label{sec:intro}

\begin{figure}[htb!]
    \centering
    \includegraphics[width=0.99\linewidth]{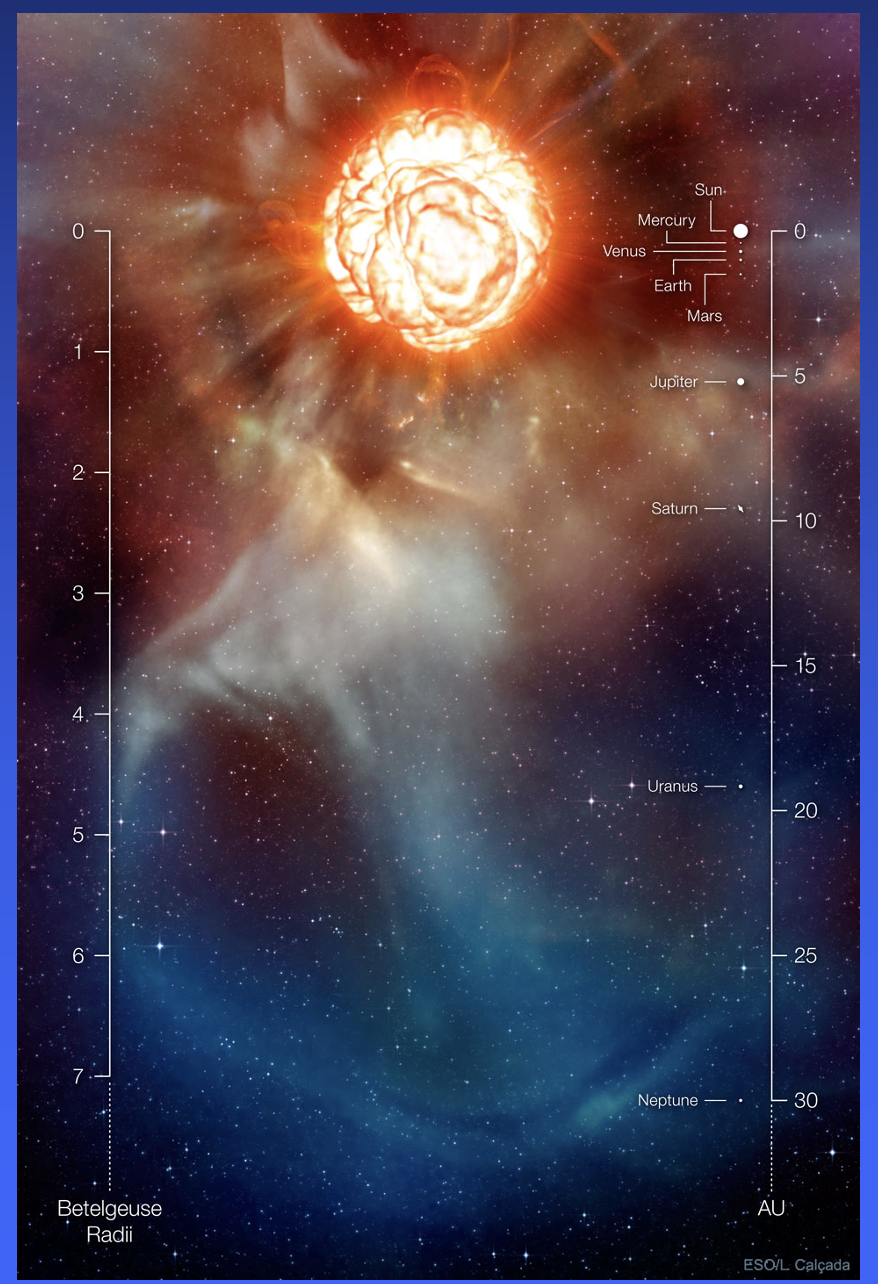}
    \caption{Schematic showing the scale of the red supergiant Betelgeuse and its circumstellar medium compared to that of the Solar System (AU = Astronomical Units). Art by L. Cal{\c c}ada, by permission of the European Southern Observatory.
}
    \label{fig:scale}
\end{figure}

Betelguese ($\alpha$ Orionis) is a nearby, massive red supergiant (RSG) that is most likely destined to explode as a classic Type IIP supernova (SN~IIP) and leave behind a neutron star. Study of Betelgeuse thus promises insight into a broad range of issues of the structure, evolution, rotation, magnetic fields, mass loss, stellar winds, circumstellar medium, dust formation, atmospheres, chromospheres, radiative transfer, nucleosynthesis, and, eventually, the explosion of massive stars. Betelgeuse is special because its propinquity allows its image to spatially be resolved. Betelgeuse also has properties such as its runaway kinematics that may be special to it. Most massive stars arise in binary systems and there are hints this may have been true for Betelgeuse despite its current apparently solo state, which seems typical of SN~IIP. Betelgeuse shows a 420-d period that is most likely a first over-tone radial pulsation mode and variance on time-scales of 2000 d that is associated with overturn of convective plumes. Then, just to keep us guessing, Betelgeuse staged the ``Great Dimming" of 2019/2020, the detailed origin of which is still debated. Figure \ref{fig:scale} gives some sense of scale of Betelgeuse.

Despite the relatively small distance from Earth, and in some sense because of it, it has been difficult to obtain tight constraints on the distance, luminosity, radius, current and Zero Age Main Sequence (ZAMS) masses, and information about the internal rotational state and associated mixing and hence on the evolutionary state of Betelgeuse and when it might explode. The best current guess is that Betelgeuse is in core helium burning and will not explode for about a hundred thousand years, but it will be a tremendous spectacle from the Planet Earth when it does.

\section{Observations}
\label{sec:obs}

Valuable summaries of basic observational properties of Betelgeuse are given by \citet{Dolan16} and \citet{Joyce20}. Here we summarize some key aspects.

\subsection{Distance}
\label{subsec:distance}

Even recently the distance to Betelgeuse has been known to only 20\% ($D \approx 197\pm45$ pc; Harper et al. 2008, 2017), a situation that was not improved by the {\sl Gaia} mission that provided accurate parallaxes but that saturates on such a bright star or is rendered less certain by transient star spots \citep{chiavassa22}. Key properties such as radius and luminosity were thus significantly uncertain, $R$ to within 20\% and $L$ to only 40\%. Estimates of mass that determine the evolution depend sensitively on $L$ and $R$ and thus also remained uncertain. The effective temperature that can be determined independent of distance has its own intrinsic uncertainties. \cite{Dolan16} estimated $T_{eff} = 3500\pm350$~K. Within these uncertainties, models of contemporary Betelgeuse could be brought into agreement with observations of $L$, $R$, and $T_{eff}$ all the way from the minimum--luminosity base of the giant branch to the tip of the red supergiant branch (RSB) \citep{Wheeler17}. Recent work has proposed ways to reduce the uncertainly in distance, but with conflicting solutions converging on either the base (\S \ref{subsec:color}) or the tip of the RSB (\S \ref{subsec:pulsation}).

\subsection{Spatial Resolution}
\label{subsec:resolution}

\begin{figure}[htb!]
    \centering
    \includegraphics[width=0.99\linewidth]{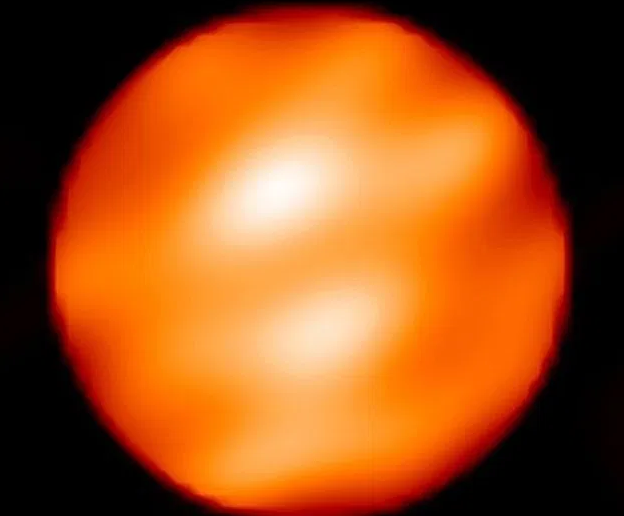}
    \caption{Spatially-resolved H band image of Betelgeuse. From \citet{haubois09} by permission of X. Haubois, ESO/Observatoire de Paris, and Astronomy \& Astrophysics.
}
    \label{fig:image}
\end{figure}

A special characteristic of Betelgeuse is that its relatively small distance allows its surface to be spatially resolved with appropriate instrumentation as shown in Figure \ref{fig:image}. The photosphere of Betelgeuse subtends an angle of $\sim 40$ milliarcseconds that can be resolved with ground-based interferometry in the optical and infrared \citep{haubois09,2016A&A...588A.130M,2022A&A...661A..91L} and submillimeter \citep{2017A&A...602L..10O,Kervella18,haubois19} or from space with the Hubble Space Telescope (HST) \citep{1996ApJ...463L..29G,1998AJ....116.2501U}.

\citet{1996ApJ...463L..29G} resolved Betelgeuse spatially by obtaining images with the HST Faint Object Camera in 10 resolution elements across the surface. They found the ultraviolet diameter of Betelgeuse to $108 \pm 4$ mas, a factor of 2.2 larger than the optical diameter, suggesting an extended chromosphere in analogy to the hot temperature inversion in the Sun. A single bright, unresolved area was 200~K hotter than the mean value. \citet{1996ApJ...463L..29G} suggested this surface inhomogeneity might be due to magnetic activity, atmospheric convection, or global pulsations that produce shock structures that heat the chromosphere. Spatially resolved spectroscopy with the Goddard High Resolution Spectrograph suggested the complicated dynamics of outflowing material in the chromosphere \citep{lobel01}.

\citet{haubois09} undertook H-band interferometry with the Infrared-Optical Telescope Array (IOTA) at the Whipple Observatory to measure the diameter ($44.28 \pm 0.15$ mas), effective temperature ($3600 \pm 66$ K), limb darkening, and bright or dark patches in the photosphere and surroundings.

\cite{2016A&A...588A.130M} did H-band interferometry on the VLT to explore mass loss driven by strong convective motions by mapping the shape of the envelope and following the structure of the wind from the photosphere out through the nearby circumstellar medium and into the interstellar medium. They detected a hot spot on the photosphere comparable in size to the radius of the star.

\cite{2017A&A...602L..10O} used submillimeter observations with the  {\it Atacama Large Millimeter Array} ({\it ALMA}) to study the free-free emission in the extended atmosphere of Betelgeuse. They found that the mean temperature at 1.3 stellar radii was 2760 K, a value that is less than both the photospheric temperature they gave as $T_{eff} = 3690$ K and the temperatures at 2 solar radii, implying an inversion of the mean temperature in the atmosphere. The emission showed evidence for inhomogeneous localized heating in the atmosphere of Betelgeuse, perhaps related to magnetic activity generated by large-scale convection.

We will return to the power of interferometry in \S \ref{sec:dimming}.

\subsection{Convection and Plumes}
\label{subsec:plumes}

The extended outer envelope of Betelgeuse engenders appreciable superadiabatic temperature gradients that lead to strong convection \citet{swarzschild75}. Both direct observations \citep{1996ApJ...463L..29G,1998AJ....116.2501U,haubois09,dupreestef13,2016A&A...588A.130M,2017A&A...602L..10O,Kervella18,haubois19,2022A&A...661A..91L} and models \citep{chiavassa10,goldberg22} indicate that the convective structure of the envelope of Betelgeuse is characterized by large plumes of upwardly rising hot material and inwardly cascading cooler material. The plumes in turn lead to hot and cold patches on the surface that are substantially large compared to the radius of the star \citep{2016A&A...588A.130M}. This leads to complications in determining basic quantities like the global effective temperature \citep{Levesque20}.

\subsection{Atmosphere, Photosphere, Chromosphere}

\label{subsec:atmos}

Driven by the irregular convective plumes, the outer layers of Betelgeuse reveal a complex atmospheric structure as the optically-thick convective envelope yields to a wavelength-dependent and position-dependent photosphere and chromosphere \citep{Bernat76, lim98,Plez02,2016A&A...588A.130M,2017A&A...602L..10O,2022A&A...661A..91L}. \citet{2017A&A...602L..10O} established a temperature inversion between the photosphere and chromosphere (\S \ref{subsec:resolution}).


\subsection{Mass and Mass Loss}
\label{subsec:massloss}

The ZAMS mass is a fundamental property that determines the evolution of a star. In the case of Betelgeuse, the uncertainty in distance and other factors yields intrinsic uncertainty in the ZAMS mass. Betelgeuse qualifies as a massive star, but estimates of the ZAMS mass vary from 10 to 25 \msun. In recent estimates, \citet{Dolan16} gave 17 - 25 \msun\ whereas \citet{Joyce20} found 18 - 21 \msun. This mass range destines Betelgeuse to succumb to iron core collapse and a likely catastrophic explosion. Direct collapse to a black hole is a remote possibility (\S \ref{sec:evolution}).

The subsequent evolution of Betelgeuse is not determined solely by its ZAMS mass, but also depends on abundances, rotation, stellar winds, the presence of a binary companion, and the possibility of a merger.

Mass loss on the main sequence is estimated to be less than 0.1 \msun, a small effect compared to other uncertainties. \citet{harper01} and \citet{lebertre12} determined the current mass loss rate to be $\sim 1 - 4\times10^{-6}$~\msunyr. \citet{Dolan16} adopted $2\times10^{-6}$~\msunyr. Estimates of the current wind velocity of Betelgeuse range from 3 to 14 \kms. \citet{Dolan16} adopted a range of $9\pm6$~\kms (their Table 5). The wind accelerates, so a single wind velocity may not be appropriate.

As for other RSGs, mass loss from Betelgeuse in its current configuration is episodic \citep{decin12,massey23}, a factor often neglected in prescriptions for mass loss rates. This variability is probably linked to the sporadic convective plumes and to the intrinsic pulsational properties. Magnetic fields may play a role (\S \ref{subsec:magnet}).

\cite{2022A&A...661A..91L} sought to understand convection and the  mechanisms that trigger mass loss by using linear spectropolarimetry of the atomic lines to provide velocity and hence depth information in addition to spatial distribution. The result was images of the photosphere of Betelgeuse that provide information about the 3D distribution of brightness in the atmosphere. The data revealed the velocity of vertical convective flows at different heights in the photosphere that showed that non-gravitational forces are present in the
photosphere of Betelgeuse that allow plasma to reach velocities close to the escape velocity. These forces may trigger mass loss and sustain large stellar winds.

\citet{humphreys22} argue that Betelgeuse gives evidence for discrete, directed clumpy outflows as suggested by circumstellar gas knots detected in the submm region
that are related to magnetic fields and surface activity. They argue that this clumpy outflow analogous to solar coronal mass ejections is a major contributor to mass loss from RSGs, including Betelgeuse.


\subsection{Molecules and Dust}
\label{subsec:dust}

Plasma recombining to gas continues to cool as it is ejected from the surface of Betelgeuse. If it gets sufficiently cool, the gas can form molecules through complex non-equilibrium chemistry. The molecules can then serve as nucleation sites where inorganic dust can form. Dust grain surfaces in turn provide an environment to form yet other molecules.

\citet{jenningssada98} discovered water in the atmosphere of Betelgeuse. \citet{tsuji00} confirmed the presence of water in data taken 35 years previously with the balloon-borne telescope Stratoscope II \citep{wolf64}. He proposed a molecular shell, a MOLsphere, in the atmosphere of Betelgeuse. \citet{perrin07} subsequently identified a geometrically thin shell between 1.31 and 1.43 $R_{star}$ with a typical temperature of 1550 K that contained H$_2$O, SiO, and Al$_2$O$_3$. \citet{Ohnaka09,Ohnaka11} spatially resolved the macroturbulent gas motion in the photosphere and MOLsphere of Betelgeuse for the first time.

Models presented by \citet{harper01} suggested that dust formed at about 33 $R_{star}$ at a temperature of $\sim 360$ K. \citet{Kervella18} argued that convective cells lead specifically to the production of molecular plumes and dusty knots in the north polar region of Betelgeuse. Related notions came to the fore during the Great Dimming of 2019/2020 (\S \ref{sec:dimming}).

\cite{haubois19} did near-IR interferometry to explore the connection between dust formation and mass loss from Betelgeuse. They found a halo of fosterite (Mg$_2$SiO$_4$) dust beginning about 0.5 $R_{star}$ above the photosphere, much lower than suggested by the models of \citet{harper01}. The height of molecule and dust formation may vary inhomogeneously over the surface of Betelgeuse.

\subsection{Surface Gravity}
\label{subsec:gravity}

The gravitational acceleration at the surface of Betelgeuse, the surface gravity, $g = G M/R^2$, provides an independent constraint on the ratio $R/M$. This quantity is determined from the analysis of line structure in the photosphere and is typically presented as the logarithm in base 10 of $g$ measured in the cgs system. \citet{lambert84} observed forbidden O I lines, vibration-rotation bands of second-overtone CO near 1.6 micron, NH bands between 3 and 4 microns, OH fundamental bands near 3 microns, and CN red lines near 8000 \AA\ and 2 microns, and employed sophisticated model atmospheres designed for supergiant stars. For Betelgeuse, \citet{lambert84} adopted $log~g = 0.0 \pm 0.3$. \citet{Lobel00} used near-UV, optical, and near-IR high-dispersion spectra analyzed with non-LTE radiative transfer calculations to obtained $log~g = -0.5$ that is somewhat less, even given the nominal uncertainties. Neither \citet{lambert84} nor \citet{Lobel00} considered the plume structure of the envelope and departures from spherical symmetry.

\citet{Neilson11} employed limb-darkening laws and grids of spherical model stellar atmospheres to determined $R/M = 82^{+13}_{-12}$ \rsun/\msun. From their best-fitted models, \citet{Dolan16} obtained $R/M = 40$~\rsun/\msun, substantially less than \citet{Neilson11}, and $log~g = -0.05$ for their Eggleton-based code and $log~g = -0.10$ with the stellar evolution code Modules for Experiments in
Stellar Astrophysics (\textsc{mesa}; \citealt{Paxton11, Paxton13, Paxton15,Paxton18}). The latter estimates for $log ~g$ are roughly consistent with \citet{lambert84} but appreciably larger than found by \citet{Lobel00}.

In principle, the effective gravity at the surface of a star is reduced by the centrifugal effects of rotation that is substantial in Betelgeuse (\S \ref{subsec:rotation}). For a 20 \msun\ model rotating at velocities typical of Betelgeuse, \citet{Wheeler17} found $log~g = +0.42$ at the luminosity minimum at the base of the RSG branch and $log~g = -0.48$ during carbon burning when the model had slowed due to envelope expansion. The former is somewhat beyond the upper limit set by \citet{lambert84} and the latter in close agreement with the determination of \citet{Lobel00}. For their models with a 16 \msun\ primary merging with a 4 \msun\ secondary, \citet{chatz20} found post-merger surface gravity for models merging at 300 and 250 \rsun\ to be $4.67 - 6.65$ cm~s$^{-2}$, corresponding to $log~g = 0.67 - 0.82$.

There are thus significant uncertainties in both observations and models of $log~g$ for Betelgeuse. Constraints on $log~g$ come into play in considering pulsational properties (\S \ref{subsec:pulsation}) and the possibility of a recent color change in Betelgeuse (\S \ref{subsec:color}).

\subsection{Rotational Velocity}
\label{subsec:rotation}

The rotation of Betelgeuse at the surface and at depth has implications for estimates of the current age, the current mass, the ZAMS mass, the current evolutionary state,  and the time to explosion.

Betelgeuse appears to have an anomalously large rotational velocity. Long slit spectroscopy across the minimally resolved disk of Betelgeuse obtained with the {\it Hubble Space Telescope} (HST) yielded an estimated surface rotational velocity $v_{\rm rot} \sin(i) \sim 5$~\kms\ at an inclination of $i \approx 20$\degree\
\citep{Dupree87,1996ApJ...463L..29G,1998AJ....116.2501U,2009A&A...504..115K}.  These data imply an equatorial rotational velocity of $\sim15$ \kms. The uncertainty in this quantity is itself uncertain.

More recent observations appear to further support this result even within the uncertainties imposed by large--scale convective motions on the star's surface. \citet{Kervella18} used {\it ALMA}  to resolve the surface velocity and determined that Betelgeuse rotates with a projected equatorial velocity of $v_{\rm eq} \sin(i) = 5.47 \pm 0.25$~\kms\ with an estimated rotation period of $36 \pm 8$ yr (see \S \ref{subsec:rotation}). They confirmed that the chromosphere is co-rotating with the star up to a radius of 1.5 times the continuum radius. They found that the position angle of the polar axis of Betelgeuse coincided with a hot spot in the {\it ALMA} data, suggesting that focused mass loss was currently taking place in the polar region. They proposed that a particularly strong convection cell was driving a focused molecular plume that could subsequently condenses into dust at a few stellar radii thus contributing to anisotropic mass loss (\S \ref{subsec:plumes}, \S \ref{subsec:massloss}, \S \ref{subsec:dust}, \S \ref{subsec:rotation}, \S \ref{sec:dimming}).

High rotation during the supergiant phase is not found in stellar evolution calculations of single massive stars (\S \ref{subsec:single}) -- including those that are rapid rotators at the Zero Age Main Sequence (ZAMS) -- nor expected by simple arguments of angular momentum conservation \citep{Wheeler17}.

Single massive stars lose a fraction of their mass and angular momentum through winds already during the main sequence (MS) phase. O stars with initial rotation velocities of $\sim 200$ \kms\ evolve through rapid mass and angular momentum losses to become much slower rotating B stars with $v \sin i \leq$~50~km~s$^{-1}$ (\citealt{rotation,2019A&A...622A..50H} and references therein). Simple analytic arguments \citep{chatz20} and stellar evolution calculations  \citep{1989A&AS...81...37C} suggest that a star rotating at $\sim$~200~km~s$^{-1}$ at the ZAMS is likely to decrease to $\leq$~50~km~s$^{-1}$ at the Terminal Age Main Sequence (TAMS). Similar estimates and detailed simulations of the evolution of massive stars, including mass and angular momentum losses from the ZAMS to the supergiant stage typically yield an upper limit to the equatorial rotational velocity of $v_{\rm eq} < 1$ \kms\ on the RSB \citep{2008A&A...489..685E,2012A&A...537A.146E,brott11a,brott11b}.  Measurements of giant and supergiant star rotation rates support this argument \citep{2017A&A...605A.111C}. \citet{Wheeler17} and \citet{chatz20} found a velocity of $\sim 0.1$ \kms\ high on the RSB.

{\it Kepler} observations of low-mass giant stars ($<3$\msun) showed 17 with rotational speeds up to $\sim$~18 times that of the Sun \citep{costa15}. It is possible that a yet unknown mechanism, perhaps transfer of angular momentum from inner regions by g-mode acoustic waves (\S \ref{subsec:seismology}), could account for this rapid rotation \citep{2014ApJ...796...17F,2018MNRAS.475..879T}, but it is not clear that even such mechanisms can account for the rotation of a massive RSG like Betelgeuse.

Taken at face value, Betelgeuse is thus rotating too rapidly by a factor $\sim$ 15 and perhaps as much as 150 compared to basic single-star models high on the RSB \citep{Wheeler17,chatz20,Joyce20}. Models of Betelgeuse on the RSB give a critical Keplerian velocity of $\sim 65$~\kms \citep{Wheeler17}; the observed rotational velocity is thus a substantial fraction of the escape velocity. Such a rotation may cause measureable oblateness that could complicate interpretation of the observations \citep{tatebe07,haubois09}.

There are concerns that the deduced rotational velocity is not correct, perhaps confused by the complex large scale convective flows at the photosphere. \citet{gray01} found a macroturbulence Gaussian dispersion $\sim 15$ \kms\ with a FWFM of $\sim \pm 50$ \kms\ consistent with many convection cells appearing on the stellar disk but with no evidence for giant convection cells. More recently, \citet{lopezpol} found characteristic upflow and downflow speeds of 22 and 10 \kms, respectively. \citet{jadlovsky23} argued that the projected rotational velocity $v_{\rm rot} \sin(i)$ is not trustworthy, as both edges of Betelgeuse seem to be moving towards Earth at a similar velocity.

An accurate measurement of the equatorial rotational velocity of Betelgeuse is important in order to constrain models.  Single--star rotating models give $v_{rot} \sim 15$ \kms\ only in a brief phase near the base of the RSB that would last for a few thousand years at most. It is conceivable that Betelgeuse might currently reside in this portion of the Hertzsprung Russell Diagram (HRD) by appropriately pushing $3\sigma$ error bars on $R$, $L$, and $T_{eff}$ \citep{Wheeler17}. The historical color changes of Betelgeuse characterized by \citet{Neu22} may demand that Betelgeuse is currently in this lower portion of the HRD where massive stars can change $T_{eff}$ on timescales of 1000 y (\S \ref{subsec:color}). This conclusion conflicts with the results from the careful study of the pulsation period given by \citet{Joyce20} that places Betelgeuse higher on the RSB (\S \ref{subsec:pulsation}).

One possibility to account for the high rotation velocity is that Betelgeuse has undergone a merger as it expanded and evolved up the RSB (\S \ref{subsec:merger}). Another pathway to form a rapidly-rotating supergiant is presented in \cite{2013ApJ...764..166D}. They propose that Case A Roche lobe overflow mass transfer from a $\sim 20$ \msun\ primary is enough to spin up a $\sim 15$ \msun\ secondary to high rotational velocity if the transfer occurs right after the TAMS before the ascent up the RSB (their Figure 2). This possibility requires considerable fine tuning of the binary evolution parameters and the timing of the onset of mass transfer.

Both the merger model and the Case A transfer model should be examined for testable observational consequences.

\subsection{Observed Abundances}
\label{subsec:abundances}

Photospheric abundances are yet another clue to the evolutionary history and state of Betelgeuse. The measured N/C (nitrogen to carbon) and N/O (nitrogen to oxygen) surface abundance ratios for Betelgeuse are 2.9 and 0.6, respectively, compared to solar values of N/C$=$0.3 and N/O$=$0.1 and the ratio $^{12}C/^{13}C$ is much lower than solar \citep{lambert84}. These ratios vary as massive stars burning hydrogen on the CNO cycle settle into CNO equilibrium, with N being produced at the expense of C and O. CN-equilibrium is achieved before an inhibiting gradient in mean molecular weight is established between the core and the envelope, so the excess N can quickly be transported to the stellar surface thus producing large N/C ratios. Full CNO-equilibrium is achieved only after significant hydrogen burning, so surface O depletion only occurs later.

The observation of enhanced nitrogen at the surface of Betelgeuse may be indicative of enhanced mixing, perhaps triggered by rotation \citep{2013EAS....60...17M}. The effects of rotational mixing are more pronounced at lower metallicity, higher ZAMS mass, and higher rotational velocity \citep{brott11a}. Rotational mixing may need to be supplemented by other effects such as binary evolution and magnetic fields to understand the abundance distributions in evolved massive stars \citep{brott11b}.

\citet{2022ApJ...927..115L} have used surface abundances to constrain the nature of Betelgeuse in terms of initial mass, rotation, and overshoot. They find the acceptable range of ZAMS masses is slightly larger for rotating models than non-rotating models, 12 - 25 \msun\ versus 15 - 24 \msun, respectively. They find that the initial rotation on the ZAMS must be restricted to 0.3 of the Keplerian velocity in order to fit the surface abundances of Betelgeuse as an RSG and find that some of their models could be in the phase of carbon burning or beyond.

The observed abundances in Betelgeuse are consistent with material that has been mixed to the surface in the first dredge-up phase when the convective hydrogen envelope penetrates the helium core \citep{lambert84,Dolan16}. This constrains Betelgeuse to have passed the base of the RSB and to be ascending the RSB, consistent with the results of \citet{Joyce20} but perhaps in contradiction with the conclusions of \citet{Neu22} (\S \ref{subsec:color}).

\subsection{Kinematics, Nearby CSM, ISM, Bowshocks}
\label{subsec:csm}

\begin{figure}[htb!]
    \centering
    \includegraphics[width=0.99\linewidth]{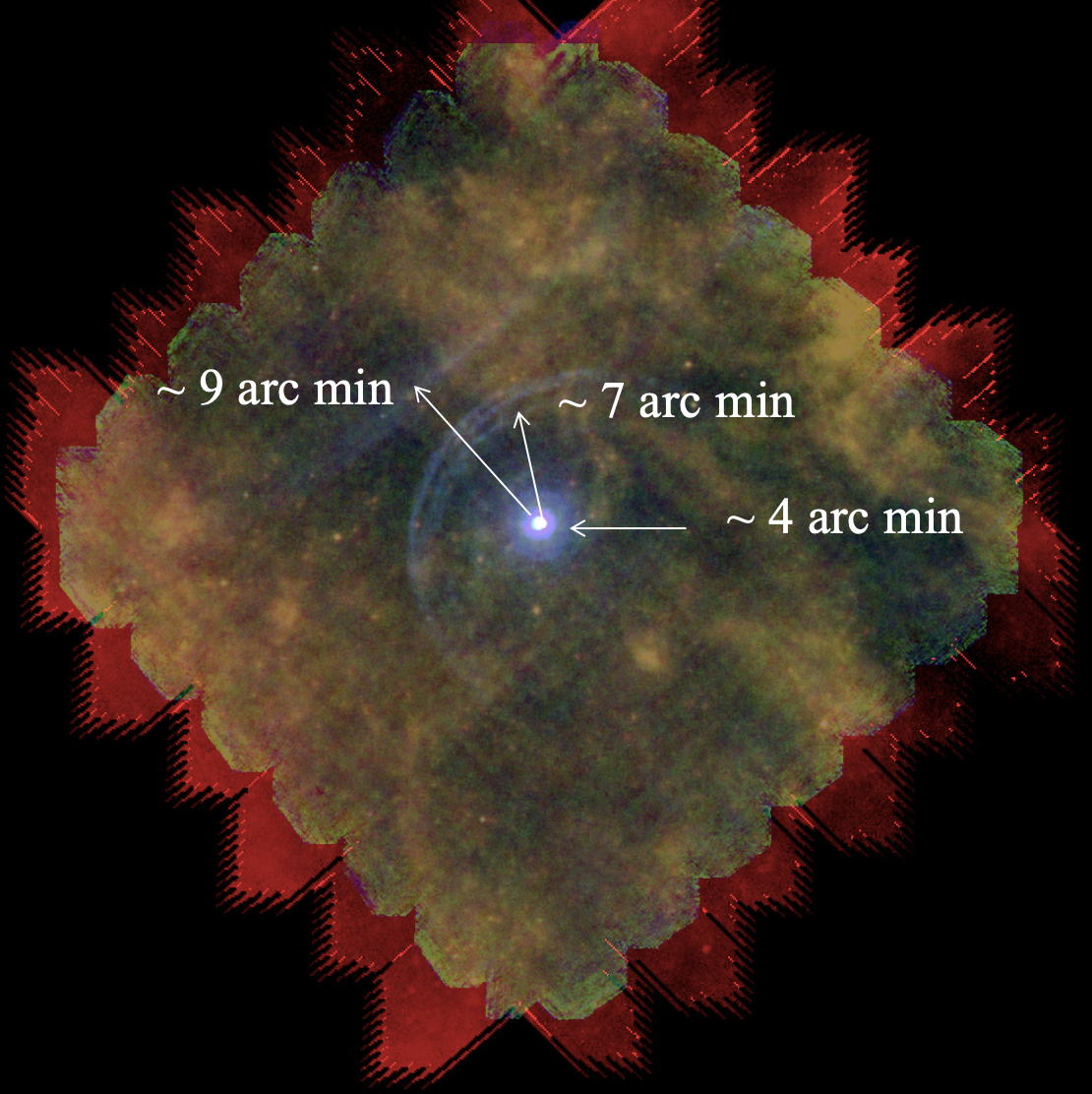}
    \caption{Structure in the large scale CSM surrounding Betelgeuse observed by the {\it Herschel} mission. Note the prominent bow shock at 7 arcmin that is in the direction of the spatial velocity of Betelgeuse. From \citet{decin12}. Adapted by permission of L. Decin and Astronomy \& Astrophysics.
}
    \label{fig:csm}
\end{figure}

In addition to perhaps being a rapid rotator, Betelgeuse is also a known runaway star with a measured space velocity of $\sim$~30~km~s$^{-1}$ and a kinematic age of $\sim$~7--11~Myr \citep{2008AJ....135.1430H,2017AJ....154...11H}.

As shown in Figure \ref{fig:csm}, the flight of Betelgeuse through the interstellar medium is also illustrated by {\it HST} and {\it Herschel} observations of a bow shock forming a swept--up shell of material of $\sim$~0.14~$M_{\odot}$ at a radius of $\sim$~6--7~arcmin corresponding to a physical distance of $\sim$~0.8~pc using a distance to Betelgeuse of $\sim$~400 pc \citep{1997AJ....114..837N,decin12} (current estimates of the distance are less by a factor of two or three; \S\S \ref{subsec:distance},\ref{subsec:pulsation}). The prominent bow shock is in the same direction as the kinematic motion, indicating a peculiar velocity with respect to the local standard of rest of $v \approx 25$ \kms\ \citep{2008AJ....135.1430H} or perhaps as much as 35 \kms\ \citep{vanloon2013}. The morphology of this structure is attributed to wind from the star sweeping up interstellar medium in the direction of motion \citep{2012A&A...541A...1M,decin12,mackey14}. The observations also show a smaller ring of material with a diameter of about 4 arcmin \citep{lebertre12}. One explanation is that this is wind mass that is radiation–impeded by external radiation \citep{mackey14}. There is also an odd, very linear feature about 9 arcmin away, beyond the bow shock, that remains unexplained \citep{1997AJ....114..837N,decin12}. \citet{Wheeler17} noted that a merger event might have some relation with the interstellar shells of higher density in the vicinity of Betelgeuse. The strangely linear feature at 9 arcmin might be related to the square axisymmetric circumstellar nebula recently discovered around the B9 Ia star HD93795 by \citet{Gvar20}. Such a connection might in turn suggest that Betelgeuse had undergone some previous mass expulsion.

Proposals to account for the high space velocity of Betelgeuse include multi--body stellar interactions in its birth cluster and the possibility that a binary companion underwent a supernova explosion \citep{blaauw61,vanloon2013}.
In a study of the 30 Doradus region of the Large Magellanic Cloud, \citet{Sana22} conclude there are two different populations of massive runaway Main Sequence O stars: a population of rapidly spinning ($v_{\rm eq} \sin(i)> 200$~\kms) but slowly moving ($v = 25 - 60$~\kms) runaway stars and a population of slowly rotating ($v_{\rm eq} \sin(i)< 200$~\kms) rapidly moving ($v > 60$~\kms) stars. They found no rapidly spinning, rapidly moving stars in their sample. \citet{Sana22} argue that slowly moving rapidly spinning stars result from binary ejections, while rapidly moving slowly spinning stars result from dynamical ejections, with slowly moving rapidly spinning stars and hence binary evolution dominating the current massive runaway star population in 30 Doradus. Betelgeuse nominally belongs in the slowly moving rapidly spinning runaway category.

Backwards extrapolation of the current trajectory of Betelgeuse has led some to suggest that its possible birthplace is the Orion OB1a association \citep{2005AJ....129..907B}. Others have argued that a backward extrapolation of its known space velocity does not appear to bring Betelgeuse close to any plausible sub--association of OB1 as its birth place \citep{2008hsf1.book..459B}. \citep{2008hsf1.book..459B} suggests a two step process: (1) a dynamical ejection of a binary within the first few million years after the formation of Betelgeuse's birth cluster, and (2) a subsequent merger of the binary or a supernova explosion of the more massive component, releasing the surviving now single Betelgeuse at some post MS stage of its evolution.

Work on the kinematic effects of supernovae in massive star binary systems tends to discourage the conjecture of the previous explosion of a companion to Betelgeuse. \citet{Renzo19} confirm that of order 20 - 50\% of massive star binaries merge rather than undergoing disruption. They also find that by far the largest fraction of binaries disrupted by the collapse and explosion of the primary result in ``walkaway" rather than ``runaway" stars. The velocity distribution of the ejected companion peaks at about 6 \kms. For secondaries more massive than 15 \msun, as likely applies to Betelgeuse, only $\sim 0.5\%$ have velocities of 30 \kms\ and above, as appropriate to Betelgeuse.

These results suggest that, while non-zero, the likelihood that the space motion of Betelgeuse resulted from the previous explosion of a companion is small. The results depend on assumptions about primordial binaries, among other things, but the general result is that it is easier to generate walkaway stars than runaway stars. A runaway binary is likely to be rare, but is not precluded.

As discussed above, a reasonable alternative is that the proper motion of Betelgeuse arises from stellar dynamics in its natal cluster \citep{poveda67,ohkroupa16,schoettler19}. Early ejection as a single star either by the disruption of a cluster binary or dynamical escape from a cluster are unlikely to yield a rapid rotator in the present supergiant stage. Even if spun up on the ZAMS, its rotation on the RSB would be slow. If a previous binary companion exploded, then it clearly could not have merged with the current Betelgeuse as discussed in \S \ref{subsec:merger}.

The origin of the space motion of Betelgeuse is thus one more fascinating open question about this tantalizing star. Whether Betelgeuse attained its proper motion from the explosion of a companion or from cluster dynamics, if it emerged as a single star then the apparent observed equatorial velocity remains an issue.

A possible way to account for both the space motion and the equatorial velocity would be to provide the space motion by cluster dynamics and ejection of a binary, of which the star we currently observe as Betelgeuse was the primary, and a subsequent merger along the RSB. This is, admittedly, an improbable string of events. \citet{ohkroupa16} find that a majority of ejected massive binaries have a period shorter than $10^5$ days. Supergiant branch merger models have a typical presumed orbital period of about 30 years or $10^4$ days \citep{Wheeler17}. Having a rather massive companion might increase the likelihood that the binary remains intact upon ejection from the natal cluster. Current results allow for that possibility.

We note that while Betelgeuse may have moved hundreds of pc during its main sequence lifetime, it is expected to have moved only $\sim 2$ pc during the 100,000 years or so it has been in core helium burning as a RSG.

\subsection{Magnetic Fields}
\label{subsec:magnet}

The atmosphere of Betelgeuse is observed to harbor magnetic fields of $\sim 1$ G as as measured by circular polarization \citep{mathias18} and as inferred from the Zeeman effect \citep{auriere10}. These fields are thought to originate from local low-scale convective activity and associated non-linear dynamo action in Betelgeuse or perhaps from giant convective cells on its surface \citep{dorch04}.

We have noted earlier that magnetic fields may play a role in localized hot spots on the surface of Betelgeuse, in the formation of the chromosphere, and in clumpy mass loss. \cite{2012MNRAS.422.1272T} addressed the effects of magnetic fields on winds, dust, and the structure of the photosphere and chromosphere of Betelgeuse.


\subsection{Pulsation Periods}
\label{subsec:pulsation}

We noted in \S \ref{subsec:gravity} that measurement of the surface gravity provides a constraint on $R/M$, given an independent measurement of $R$. Stellar pulsation modes also depend on gravity, giving yet another constraint on $R/M$. Betelgeuse displays a range of periodic behavior.

Studies of Betelgeuse have long revealed a variety of pulsation modes. Of particular value is the record of optical photometry compiled by amateurs and professionals for nearly 100 years and recorded by the American Association of Variable Star Observers (AAVSO). These data reveal at least two different timescales, $\sim 388$ d and a ``long secondary period" (LSP) of $\sim 2050$ d (5.6 yr) \citep{ kiss06, chatys19}. The LSP might be related to the rotation, but the rotation period is apparently significantly longer (\S \ref{subsec:rotation}).

\citet{Joyce20} analyzed the most recent $\sim 40$ years of data from the AAVSO complemented with data incidentally produced by Solar Magnetic Ejection Imager (SMEI) observations. They find periods of $185\pm13.5$ d, $416\pm24$ d, and $2365\pm10$ d, cautioning that these periods could evolve with time. U-band observations are relatively rare. \citet{ogane22} presented 23 years of UBVRI data obtained at the private Ogane Hikari Observatory and found periods of $\sim 405$ d and $2160$ d.  \citep{jadlovsky23} presented an analysis of spectroscopic and photometric variability in the UV and optical, finding photometric periods of $417\pm17$ d and $2190\pm270$ d and radial velocity periods from spectroscopy of $415\pm11$ d and $2510\pm440$ d. The radial velocity determined from ultraviolet spectra show longer periods of variability that may be related to the outflowing wind.

Models of RSGs show that pressure-mode or p-mode radial pulsations can be driven by the opacity or $\kappa$-mechanism in the hydrogen ionization zone. In this mechanism, the opacity varies out of phase with the luminosity, being lower when the star is compressed and hot releasing radiant energy and allowing more compression and higher when the star expands and cools thus blocking radiant energy and driving more expansion. Simulations yield mass-dependent periods of the fundamental of years in both linear and nonlinear models \citep{ligong94,heger97,yooncant10,Paxton13,Dolan16,goldberg22}. Modeling pulsation processes may require 3D, time-dependent convection, or otherwise more sophisticated physical formalisms that are beyond the scope of
typical 1D stellar evolution programs, but 1D analyses already provide useful insights.

\citet{Joyce20} used 1D hydrodynamical models and the GYRE pulsation module of \textsc{mesa} to analyze the pulsations and provide new constraints on $R/M$ for Betelgeuse. They deduced that the 416 day period represents oscillation in the fundamental mode, driven by the opacity mechanism, and that the 186 day period represents the frequency of the first  overtone of radial pulsations. \citet{Joyce20} also used the period information to provide a tighter constraint on the radius of Betelgeuse, $R = 750^{+62}_{-30}$~\rsun~($3\sigma$), compared to the previous estimate of 887~\rsun. Surprisingly, this led to a tighter constraint on the distance and parallax than previous methods, $D = 165^{+16}_{-8}$ pc with <10\% uncertainty compared to the previous estimate of 197 pc with 20\% uncertainty, and tighter constraint on the ZAMS mass, 18 - 21~\msun\ and the current mass, 16.5 – 19~\msun. \citet{Joyce20} do not give a surface gravity to compare with atmospheric observations (\S \ref{subsec:gravity}). They give an extensive discussion of model degeneracies that make estimates of L and $T_{eff}$ uncertain.

 With the new constraints, \citet{Joyce20} concluded that Betelgeuse is in core helium burning, with $\sim$ 100,000 years to go before explosion.

\subsection{Recent Change in Color?}
\label{subsec:color}

Another approach to determining the mass, luminosity, radius, distance, effective temperature, age, and current evolutionary state of Betelgeuse is to study the color evolution from the historical record. In a recent rigorous analysis of extensive multi-cultural historical literature including Tycho Brahe's comparison of Betelgeuse to his supernova of 1572 and to Aldebaran, \citet{Neu22} [and new summary in Astronomy and Geophysics] argue that Betelgeuse has significantly changed color over the last two millennia.

Contemporary Betelgeuse is, as can be verified by casual naked eye observation, red, with a formal color of $B - V = 1.78\pm0.05$ mag. \citet{Neu22} argue that 2000 years ago Hyginus in Rome reported Betelgeuse to have a color similar to Saturn that is equivalent to $B - V = 1.09\pm0.16$ mag and that Sima Qian in China independently reported Betelgeuse to be ``yellow," a condition that \citet{Neu22} quantify to be $B - V = 0.95\pm0.35$ mag. \citet{Neu22} estimate that these historical estimates of color differ from the contemporary color by $5.1\sigma$. In contrast, Antares has always been reported as red for over 3000 yr.

Taken at face value, this color change of Betelgeuse represents a strong constraint on evolutionary models.
\citet{Neu22} compare their estimates of historical and contemporary colors of Betelgeuse to the \textsc{mesa} Isochrones and Stellar Tracks (MIST) of \citet{choi16}. They deduce that Betelgeuse is likely to be near the cool end of the Herzsprung Gap and less than 1000 yr past the minimum of the RSB when relatively rapid changes in color are expected. \citet{Neu22} specifically argue that the color evolution and location in the color-magnitude diagram constrain the ZAMS mass to be $\sim14$~\msun\ with a current age of $\sim14$~Myr. This deduction is in distinct contrast with the location in the Hertzsprung-Russell Diagram, the ZAMS mass (18 - 21~\msun), and the evolutionary state deduced by \citet{Joyce20} .

In their study of the rotation of Betelgeuse, \citet{Wheeler17} noted that the radius increases and the surface velocity plummets as models proceed across the Hertzsprung gap and up the RSB. The only position in the Hertzsprung-Russell Diagram for which single star models could plausibly give the observed equatorial rotation of $\sim15$~\kms\ (\S \ref{subsec:rotation}) is when the models first approach the base of the red supergiant branch (RSB), having crossed the Hertzsprung gap but not yet having ascended the RSB. This condition is similar to that deduced by \citet{Neu22}. \citet{Wheeler17} argued that because that phase is so short ($\sim100$~yr), that possibility is highly unlikely. Rather, they suggested, Betelgeuse may have been in a binary system that merged (\S \ref{subsec:merger}), producing the observed rotation near the upper tip of the RSB, the condition deduced by \citet{Joyce20}. If \citet{Neu22} are correct in their interpretation of the historical data, their results are a challenge to models, including merger models, that attempt to place contemporary Betelgeuse in the upper reaches of the RSB.

At this writing, the conflict between \citet{Joyce20} and \citet{Neu22} is unresolved. \citet{Wheeler17} noted that a solution near the base of the RGB, as advocated by \citet{Neu22}, would yield an excessively large surface gravity, $log~g \approx +0.42$ (\S \ref{subsec:gravity}). This may mitigate against the solution of \citet{Neu22}, but a proper resolution would involve identifying a flaw in either the analysis of \citet{Joyce20} or that of \citet{Neu22}.

Once again, an important factor is the distance. \citet{Neu22} favor a distance of $151.5\pm19$~pc as determined from Hipparchos data \citep{vanleeuwen07} rather than greater distance of $197\pm45$~pc determined by \citet{2008AJ....135.1430H}, for which they consider the {\it ALMA} distance less certain. With the larger distance, \citet{Neu22} find a ZAMS mass of 17 or 18 \msun, closer to the result of \cite{Joyce20}. On the other hand, Joyce et al. (2020) favor a distance of $\sim 165$ pc, closer to the preferred value of \citet{Neu22} despite their other disagreements. More accurate determinations of the surface gravity by spectral analysis and modeling would also be useful.

Given the uncertainties, it is possible that \citet{Neu22} and \citet{Joyce20} could be brought into agreement in terms of ZAMS mass, L, and $T_{eff}$ but still disagree on the corresponding evolutionary state, near the end of the Hertzsprung Gap, or substantially up the RSB.

Another possibility is that other surface activity analogous to the recent Great Dimming (\S \ref{sec:dimming}) may have caused color changes. It would be interesting if such a possibility could be ruled out.


\subsection{Asteroseismology}
\label{subsec:seismology}

Section \ref{subsec:pulsation} dealt with the fundamental global pulsation properties. There could, in principle, be other temporal signals coming from the depths of Betelgeuse that give yet more evidence of the inner structure and evolution, perhaps of unorthodox evolution such as a merger.

Over the last two decades there has been tremendous progress in using the technique of asteroseismology to explore the depths of stars from the Sun to evolved giants. High precision $\mu$-magnitude space--based photometry from the $\it CoRoT$ and {\it Kepler} missions showed complex but interpretable variations due to acoustic signals arising from deep within stars that is analogous to exploring the core of the Earth with seismic signals \citep{aerts10}. Study of these signals revealed the inner rotation of the Sun and understanding of the structure, rotation, and inner magnetic field distribution of thousands of stars from the ZAMS to the red giant branch, especially those of low mass that are technically easier to analyze.

The question then arises as to whether such asteroseismology techniques can be applied to Betelgeuse and other RSB stars. The potential is great. The inner structure of evolved massive stars is suspected to yield complex convective regions that will generate acoustic signals in the form of pressure modes and gravity waves. These should get especially intense late in the evolution near core collapse when the convective timescales become comparable to the nuclear burning timescales \citep{arnett11,couch15,chatz16}. Convective regions should hammer on the inside of the star with increasing violence and decreasing timescale as the star nears core collapse. In practice, it is difficult to do asteroseismology of RSB stars because typical oscillation periods are long and because the oscillations are affected by complex processes in the atmosphere and wind (\S\S \ref{subsec:plumes}, \ref{subsec:atmos}, and \ref{subsec:massloss}) that affect the boundary conditions employed in the analysis but that are not well understood \citep{aerts15}. In addition, Betelgeuse is too bright to study with traditional telescopes on the ground or in space due to instrument saturation.

In principle, asteroseismology could be used to determine the evolutionary stage of Betelgeuse since interior acoustic activity should get more intense with time and carry signals specific to certain stages of evolution, especially oxygen and silicon burning in the years or days before core collapse. The added mass and angular momentum and associated plumes and mixing might give evidence of a merger (\S \ref{subsec:merger}). The key question is whether some of that acoustic power reaches the surface. Could one see small perturbations on the surface of Betelgeuse given the extensive convective envelope?

The potential of asteroseismology to glean an understanding of the interior structure of Betelgeuse in particular and RSG in general has been explored theoretically.  Following \citet{2014ApJ...780...96S}, detailed stellar models can be used to estimate characteristic acoustic frequencies driven by inner convection as $\omega = v_{conv}/H_p$, where $v_{conv}$ is a convective velocity and $H_p$ is an appropriate scale height associated with a given convective region. The outer extended convective envelope has a characteristic cutoff frequency, $\omega_{cut} = c_s/H_p$, where $c_s$ is the sound speed, below which any acoustic signal cannot effectively propagate.

Typical signals from late in the evolution are potentially observable, but the envelope cutoff, propagation efficiency, wave effervescence, damping, and shock dissipation probably muffle all the inner convective noise \citep{Fuller17,romatzner17,Nance18}. The largest envelope pressure waves may arise from wave heating during core neon burning and a third carbon shell burning phase a few years before core collapse because later, more intense waves associated with oxygen and silicon burning  do not have time to reach the surface before core collapse \citep{Fuller17}. The shock dissipation of the acoustic luminosity generated in the very late stages of burning may eject some mass into the CSM \citep{Fuller17,romatzner17,morozova20}.

Most of the work on the issues described here have been done with spherically-symmetric codes, albeit ones that can treat angular momentum and its transport. Some work has been done in 2D \citep{leung20}, but a more complete understanding probably requires 3D studies \citep{tsang22}. Beside effects on late-time ejection of mass from the extended envelope, effective 3D porosity of the envelope may mitigate some of the wave damping effects and allow some asteroseismological signals to percolate to the surface causing diagnostic brightness variations even at earlier evolutionary phases.

\section{Evolutionary Models}
\label{sec:evolution}

\subsection{Single Star Models}
\label{subsec:single}

The evolution of single massive stars, both non-rotating and rotating, has been discussed extensively in the literature \citep{brott11a,brott11b,ekstrom12,snex,Wheeler17,sukh18,chatz20} 
Models of these stars show that hydrogen is burned on the CNO cycle in a convective core yielding a helium core of about 1/3 the original ZAMS mass. The helium core contracts and heats, and a thin hydrogen-burning shells forms at its surface. The shell sits at a node in the structure such that as the core contracts, the outer envelope expands becoming large in radius and convective.

Helium eventually ignites in the center, forming a core of carbon and oxygen. Contraction of this core results first in carbon burning and then the burning of heavier elements as the inner core contracts and heats. Shells form burning helium, carbon and other elements. Convection in these shells is expected to produce intense acoustic waves (\S \ref{subsec:seismology}).

Near the end of the star's lifetime, a core of silicon forms. Burning of silicon yields a core of iron. Iron is endothermic in terms of its thermonuclear properties. Within days of the formation of the iron core, it will absorb thermal energy from the star, reduce the pressure, and trigger catastrophic dynamical collapse to form a neutron star, or perhaps a black hole.

For the case of a neutron star, most likely for Betelgeuse, most of the kinetic energy of collapse will be lost to neutrinos but of order 1\% will be deposited in the inner regions, sufficient to cause a violent explosion of the star, ejecting the outer layers, and leaving behind the neutron star (\S \ref{sec:explo}).

\subsection{Common Envelope Evolution}
\label{subsec:cee}


It has been well established that a majority of O and B stars are in binary systems \citep{Sana12, deMink14, Dunstall15, costa15, 2019A&A...624A..66R, 2019A&A...631A...5Z}, so it is {\sl a priori} likely that Betelgeuse began as a binary system. The implication is that many RSG -- including Betelgeuse -- that appear to be single now have undergone mergers.

An important implication of the potential that Betelgeuse arose in a binary system is that Betelgeuse may have undergone common envelope evolution (CEE) sometime during its history. CEE is expected when the two stars in a binary are sufficiently close they interact as the more massive star evolves, expands, fills its Roche Lobe, and transfers mass to its lower-mass companion. In some circumstances, the companion cannot ingest the transferred material as rapidly as the primary loses it, and the excess mass forms a red giant like envelope surrounding the secondary and the evolving core of the primary. The secondary orbiting within the common envelope (CE) will undergo drag and spiral inward toward the evolved core. While the details are complex, there is then a potential for the secondary to merge with the core of the primary (\S \ref{subsec:merger}). The result could appear to be a single star, but with an inner structure rather different than would be expected of a single star of the same luminosity, radius, and $T_{eff}$.

CEE can result in several types of anomalous mixing within the core and between the core and the surface of the star. The inspiral phase leads to increased equatorial rotation and thus chemical mixing via rotational mechanisms. Plume mixing and nucleosynthesis occur during the moment of the final tidal disruption of the secondary, and merger with the core of the primary will affect the structure of the material inside and around the core of the primary. Details depend on whether the plume mixing is strong enough to rejuvenate hydrogen burning in the core. On longer timescales, rotational mixing can dredge some $\alpha$-enhanced material from the inner regions to the surface (\S \ref{subsec:abundances}).

\citet{Ivanova16} (see also
\citealt{MorrisPod07, Taam10, Ivanova13, IvanovaJP, MacLeod18, chatz20, 2022arXiv221207308R}) describe the basic phases of CEE and the mechanisms for treating it in 3D and 1D. There are three stages to the process, each with associated loss of mass and angular momentum: 1) a precursor phase when the stars begin to interact and co-rotation is lost; 2) a plunge-in phase with a large rate of change of orbital separation and a timescale close to dynamical, at the end of which most of the mass of the CE is beyond the orbit of the companion; and 3) a self-regulated slow inspiral of the companion. There are two basic endpoints to CEE: formation of a compact binary system and merger. For mergers, \citet{IvanovaPod03} differentiate three outcomes: a quiet merger, a moderate merger, and an explosive merger. Only the former leaves behind an RSG and hence is pertinent to Betelgeuse.

An important aspect of the problem is the deposition of the mass and orbital angular momentum of the secondary. In 3D simulations most of the initial angular momentum of the secondary is deposited in the outer layers of the primary envelope. Mass and angular momentum are lost by dynamical interaction, outflow driven by recombination, and shrinking of the orbit. The surface layers are ``shock heated" and quickly ejected prior to the plunge-in \citep{zhaofuller20}. The slow inspiral often begins with an envelope that is significantly reduced in mass and angular momentum. In some cases, recombination outflow can eject nearly all the envelope during the slow inspiral. The exception to these cases of extreme mass loss is when the primary is substantially more massive than the secondary. For small secondary masses, the fraction of mass lost in the precursor phase and the plunge-in phase is of order $q$, the mass ratio of secondary to primary.

In their treatment of a red giant of modest mass (1.8 \msun), \citet{Ivanova16} find that companions of mass less than 0.10 \msun, corresponding to about 5\% of the primary mass, undergo merger. The time to merger is about 1000 d, long compared to the dynamical time of the CE but short compared to the thermal or evolutionary time of the primary. While these results do not necessarily scale with mass, this suggests that for many cases of interest here, a companion of about 1 \msun\ undergoing CEE with a primary of about 20 \msun\ is likely to quickly undergo merger while sustaining a substantial envelope, as Betelgeuse is observed to have.

The plunge-in phase is expected to induce very asymmetric structures and the slow inspiral to yield appreciable departures from spherical symmetry that can be simulated in 3D but are beyond the capacity of 1D models. In 3D there is a significant density inversion in the vicinity of the companion and rather little material near the center of mass of the binary. On the other hand, the 3D simulations often treat the companion star and the red giant core as point sources. In 1D, the primary core, at least, can be modeled in more detail. A 1D code like \textsc{mesa} conserves energy and angular momentum within expected numerical accuracy. \textsc{mesa} also automatically handles energy released by recombination as the envelope expands and the angular momentum is lost in winds. In some 1D simulations of CEE, the companion is treated in a ``thin shell" approximation.

\citet{chatz20} argue that for massive primaries with mass ratios $q~< q_{\rm blue}$ (where 0.25$~< q_{\rm blue} <~$0.33) and initial period, $P_{\rm i}$, greater than a few tens of days, mass transfer starts in early Case B mass transfer. This situation arises when hydrogen is exhausted in the primary, so the primary has evolved off the main--sequence but not yet ignited helium, and while the secondary is still on the main-sequence. This mass transfer is rapid and results in the primary envelope engulfing the much-lower mass secondary. The secondary spirals inward producing a merger.

\begin{figure}[htb!]
    \centering
    \includegraphics[width=0.99\linewidth]{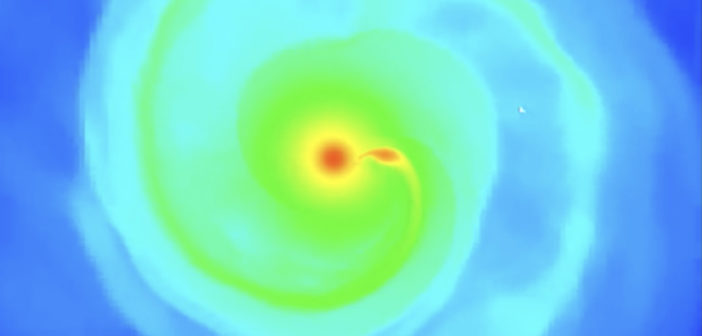}
    \caption{Density profile from a 2D simulation of a 16~\msun+1~\msun\ merger occurring at an initial separation of 12~\rsun\ showing the formation of the tidal stream (teardrop shape to the right of center) within the common envelope (outer green and blue green) as the secondary fills its Roche Lobe and is disrupted by the core of the primary (large red dot in the center). From \citet{chatz20}.
}
    \label{fig:plume}
\end{figure}

In this scenario, the helium core of the primary is surrounded by a H--burning shell. When the secondary reaches the critical tidal disruption distance from the core of the primary, a tidal stream will form that transports fresh H fuel toward the core \citep{2002PhDT........25I,2002MNRAS.334..819I,2003fthp.conf...19I} as shown in Figure \ref{fig:plume}. Mixing can thus happen if the mass transfer stream can penetrate the core \citep{IPS02}. The depth of penetration of the stream into the core depends on the direction, entropy, width, and angular momentum of the stream, the rotation, orientation, and mass of the secondary, on the density structure and relative rotation of the core, and on fluid instabilities.

The penetration depth of the stream into the core of the primary determines the extent of its rejuvenation; if fresh fuel reaches the core then core H--burning will be re--ignited and the star may evolve toward the blue supergiant (BSG) phase. If, on the contrary, the stream does not penetrate deeply into the core but rather converges with the H--burning shell, then the star will continue to evolve toward the RSG stage. \citet{chatz20} confirm, by using the arguments presented in \citet{2002MNRAS.334..819I}, that none of the models they explored (secondaries in the range 1--4~$M_{\odot}$ merging with primaries in the range 15--17~$M_{\odot}$) undergo stream--core penetration.

These results suggest that the ``quiet merger" described above is more relevant to the case of Betelgeuse. In that case, the amount of orbital angular momentum depends mostly on the binary separation when the primary overflows its Roche lobe. The total angular momentum deposited in the envelope of the primary depends on the radius of the primary when it engulfs the secondary during its crossing of the Hertzsprung gap.

\subsection{Merger Models}
\label{subsec:merger}

Of primary interest for Betelgeuse is how and under what circumstances a merged system could end up rotating at $\sim 23$\% of the critical velocity, as observations suggest (\S \ref{subsec:rotation}). Merger models provide a reasonable ``natural" explanation for why Betelgeuse has a large, but sub-Keplerian equatorial velocity \citep{Wheeler17, chatz20, Sull20}. These results do not prove, but do allow that Betelgeuse might have merged with a lower mass companion. Betelgeuse might look substantially the same whether it merged with a 1 or 10 \msun\ companion. \citet{Joyce20} concluded that Betelgeuse merged prior to the later carbon-burning phases, but see \citet{2022ApJ...927..115L}.

While the hypothesis that Betelgeuse might have merged with a companion is credible and consistent with the {\sl a priori} estimate that Betelgeuse has a probability of $\sim 20$\% of being born in a binary system \citep{deMink14}, it raises a number of interesting issues involving common envelope evolution, the fate of the companion and its angular momentum, and effect on the post-merger structure of the primary.

The luminosity of an evolved massive star is typically a function of the mass of the helium core and rather independent of the mass of the envelope. If a companion merged with the core of Betelgeuse, then the current luminosity may be a measure of the core mass ($\sim$ 5 to 6~\msun), but the mass of the envelope would be rather unconstrained and probably smaller than the estimates given based on single--star models that attempt to reproduce the luminosity, radius and effective temperature. If there were a coalescence, there would be some mass ejected.

The mass lost from the system during the merger may be substantial. The 3D 16$M_{\odot}$+4$M_{\odot}$ merger model of \citet{chatz20} lost 0.5 \msun. This model accounted for rotation, but not radiative effects nor recombination. \citet{Sull20} found up to 5 \msun\ lost. The mass loss is a combination of the loss of mass accreted from the secondary plus loss of mass from the primary itself. The latter is due to winds prior to the accretion event and then the rotationally-induced mass loss after the accretion.

A main sequence companion of about a solar mass would have a mean density of about 1 \gcm3. That density is characteristic of the base of the hydrogen envelope in the RSG models, implying that a companion might not be dissolved until it reaches the edge of the helium core (see discussion of common envelope evolution and plume penetration in \S \ref{subsec:cee}). If the companion merged with the core, the evolution of the primary might be severely altered by anomalous burning and mixing effects, and surface abundances might be affected.

\citet{Sull20} used the \textsc{mesa} code to study the merger problem in a rudimentary way that nevertheless gave some insights to the relevant physical processes. They did not attempt to treat the companion as a corporeal entity, but allowed for its effects by adding the relevant mass and associated angular momentum to the outer envelope of the primary, a computational process identified as ``accretion" to distinguish it from the more complex behavior of a true merger. The HRD of all the models of \citet{Sull20} were qualitatively similar. The accretion events resulted in irregular transient loci before settling down to a rather normal evolution up the RSB to the point of collapse of the models. The models suggest that the rotation of Betelgeuse could be consistent with a primary of ZAMS mass somewhat less than 15 \msun\ accreting between 1 and 10 \msun\ in the core helium burning and core carbon burning epochs. The observed equatorial velocity might also be attained by accreting a broad range of masses onto a primary of ZAMS mass somewhat more than 20 \msun\ in the later carbon shell burning epoch.

\citet{chatz20} used the \textsc{mesa} code to compute the 1D rotating post--merger evolution of systems with mass ratio 0.06~$ < q <$0.25 that suffer an early Case B merger. In this case, unstable mass transfer occurs during during the crossing of the Hertzsprung gap. A ``stellar engineering" approach was adopted by incorporating a perturbation term that captures the effects on the specific angular momentum and entropy. This term was used to re--adjust the post--merger structure of the envelope of the primary star during the in-spiral prior to the dynamic disruption of the secondary around the He core of the primary. The magnitude of the perturbation applied is proportional to $q$ and the structure of the primary (their Equation 13). In their \textsc{mesa} simulations, the mass of the secondary was added to the core plus hydrogen-burning shell. The composition was not adjusted as done by \citet{MenonHeger17} in their models of SN~1987A (\S \ref{subsec:87a}). Post-merger profiles were computed for different primary radii corresponding to the time when the envelope of the primary engulfed the secondary (200-700~$R_{\odot}$). The initial primary radii represented initial separations corresponding to binding energies that enabled the binary progenitor system to survive a possible past ejection from its birth cluster, to be consistent with the borderline ``runaway" nature of Betelgeuse (\S \ref{subsec:csm}). The resulting models were used to investigate the rotation rate of the post--merger object.

The models explored by \citet{chatz20} were able to reproduce the overall observed properties of Betelgeuse, including its position in the HRD, its surface rotation rate, and its surface abundances, especially the observed overabundance of nitrogen (\S \ref{subsec:abundances}). Their 16$M_{\odot}$+4$M_{\odot}$ merger occurring at $\sim$~200-300$~R_{\odot}$ yielded the best fit. These models had a sustained high equatorial rotation for a few hundred thousand years after the merger.

\citet{chatz20} also presented a 3D simulation of the merger between a 16$M_{\odot}$ primary and a 1$M_{\odot}$ secondary that occurred when the primary reached a radius of $\sim$12$R_{\odot}$, right after the end of its TAMS. The simulation was performed with the 3D {\it OctoTiger} Adaptive Mesh Refinement (AMR) hydrodynamics code developed by the LSU Center for Computation and Technology (CCT) \citep{marcello2021}. Post--processing of the 3D internal structure of the post--merger object confirmed that the envelope of the primary was spun--up by a significant amount during the in-spiral phase.

The degree of envelope spin--up is, however, proportional to the primary's radius at the onset of the merger. Three--dimensional simulations of mergers occurring at larger primary radii are needed to compute post--merger structures that evolve to become rapidly--rotating supergiants. The limitation in simulating the CEE evolution of such systems in 3D is purely of computational nature; the in-spiral timescale for a 15$M_{\odot}$+1$M_{\odot}$ merger occurring at $\sim$~300$R_{\odot}$ is $\sim$1000 years, requiring a very long simulation time. In addition, the density contrast between the compact secondary and the low-density outer regions of the envelope of the primary as well as the large simulation box that would be required to include the entire system makes it difficult to adequately resolve the full structure of the secondary, its tidal disruption plume, and the dense core of the primary, requiring billions of zones rendering such calculations prohibitively expensive.

Despite these computational challenges, there are on-going efforts involving the use of point masses to represent the secondary and the core of the primary. The merger can be accelerated by the removal of a constant, yet small, amount of angular momentum per orbit. This allows the long--term evolutionary calculation of post--merger angular momentum profiles for the primary.

\begin{figure}
\centering
\includegraphics[width=3.5in]{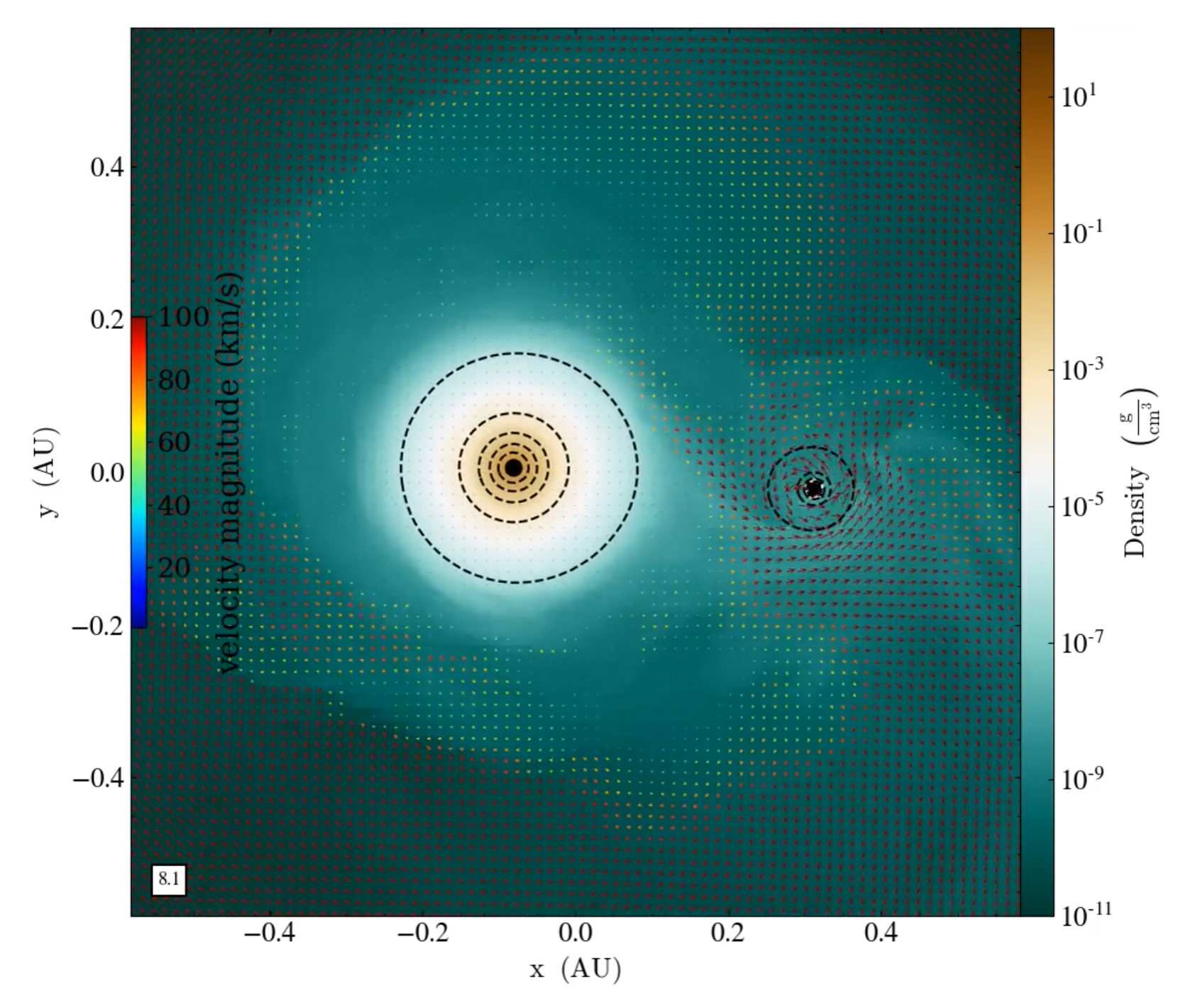}
\includegraphics[width=3.5in]{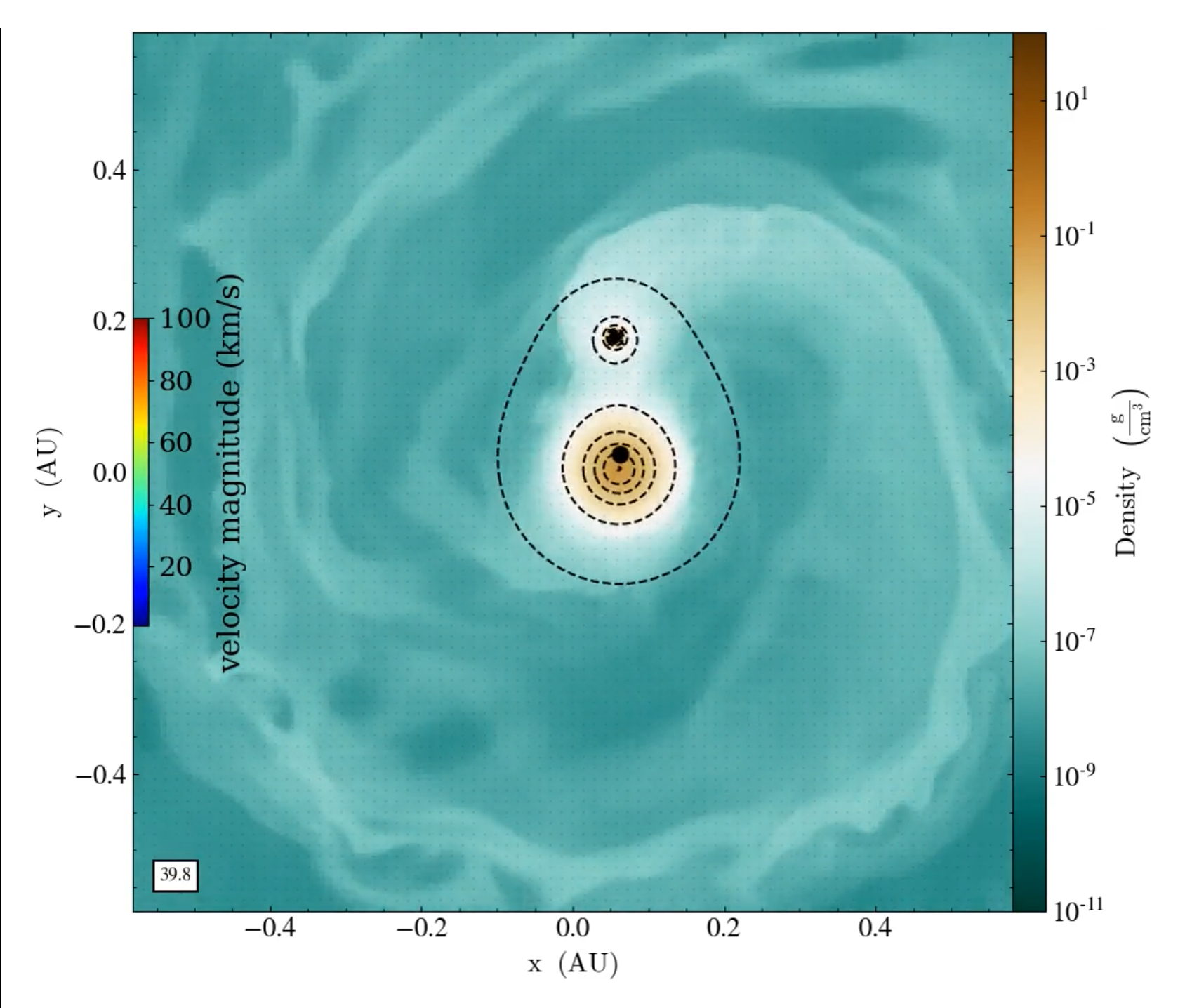}
\caption{The initial ({\it upper panel}) and final (immediately prior to the merger of the secondary with the core of the primary ({\it lower panel}) density structure of a 15$M_{\odot}$+4$M_{\odot}$ system with an initial core/secondary separation of 50~$R_{\odot}$ simulated with the 3D AMR {\it OctoTiger} code. The core of the primary and the secondary are treated as point masses. The color bar on the right represents density in \gcm3. Dashed lines represent equipotential surfaces. In the upper panel, arrows with length proportional to magnitude represent the velocity field. Velocities beyond the primary core and secondary are typically 50 \kms.
The numbers 8.1 and 39.8 in the lower left corners represent the ``orbit number" that was used in {\it OctoTiger} (even after the merger) to represent the time/phase of the simulation. From Chatzopoulos et al. 2023, in preparation.}
\label{fig:3Dmerger}
\end{figure}

An example of such a simulation involving the merger between a 15$M_{\odot}$ primary and a 4$M_{\odot}$ secondary initiated with a separation between the secondary and the core of the primary of 50~$R_{\odot}$ is shown in Figure \ref{fig:3Dmerger}. This model lost $\sim 0.4$~\msun\ in the ``mergeburst" \citep{soker06} phase right after the merger when the surface equatorial velocity was $\sim$~60~\kms. The simulation focused on the angular momentum of the remaining bound object and did not quantify the amount of angular momentum lost.

\begin{figure}[htb!]
    \centering
    \includegraphics[width=0.99\linewidth]{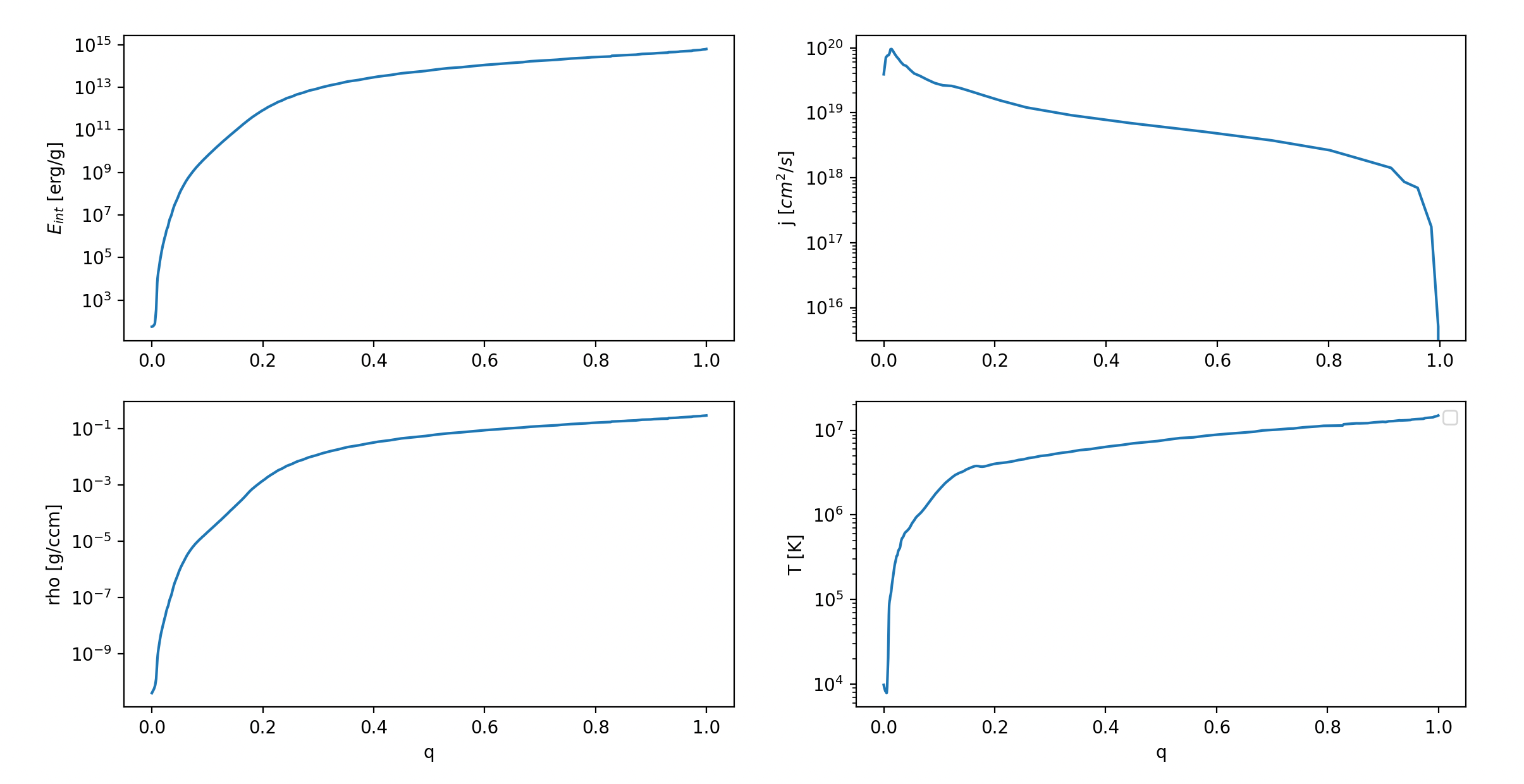}
    \caption{Energy density, specific angular momentum, density, and temperature profiles of the post--merger object following the merger of a 15$M_{\odot}$+4$M_{\odot}$ system occurring at 50~$R_{\odot}$. The original 3D profiles have been mass--weighted and angle-averaged. From Chatzopoulos et al. 2023, in preparation.
}
    \label{fig:postmerger}
\end{figure}

The spherical, mass-weighted, angle--averaged profiles for internal energy, density, and temperature and the cylindrical mass-weighted profile for specific angular momentum of the post--merger object resulting from this simulation are shown in Figure~\ref{fig:postmerger}. Note that the x-axis is q, the normalized mass-coordinate variable. The specific angular momentum, j, decreases with q, but increases in radius, so the bulk of the structure is dynamically stable. A minor decrease/instability develops in the very outer regions that contain very little mass. That behavior is in agreement with \citet{Ivanova16}.

This simulation was long and expensive. It  used a sufficiently large box to follow the unbound material after the merger for as long as possible in order to characterize the mergeburst transient and also with sufficient zones to resolve enough of the secondary structure and the primary core such that they are dynamically stable in the grid. This balance did not allow sufficient resolution to resolve the stream-core interaction in detail but such simulations will be done in the future. These proposed simulations will allow a comparison to the 3D models of \citet{Ivanova16} for lower mass systems. For related work on the
effect of radiation pressure and recombination energy in the ejection of mass in the CEE of an RSG star see \citet{lau22}.

\subsubsection{Insensitivity of Final Equatorial Velocity to Accreted Mass }
\label{subsec:finalv}

The original motivation of \citet{Wheeler17} for hypothesizing that Betelgeuse might have merged with a companion was the difficulty of accounting for the nominal currently-observed equatorial rotation velocity, $\sim15$ \kms, allowing for inclination. A companion mass of $\sim 1$ \msun\ was estimated from simple arguments based on conservation of angular momentum. Subsequent work showed that, broadly, the final rotational velocities of the models were rather independent of the companion mass accreted.

Although the treatment of the post-merger system by \citet{Sull20} and by \citet{chatz20} is rather different, the results for the final equatorial rotational velocity are very similar. This gives confidence that this quantity is somewhat robust against the details of the merger process and depends primarily on a global quantity such as the pre-merger orbital angular momentum.

If a merger occurred in Betelgeuse, the product must have settled into a state for which the rotation is sub-Keplerian. This global criterion is independent of the masses of the primary and secondary involved in the merger. The implication is that the loss of mass and angular momentum must adjust to meet this criterion rather independently of the masses involved and the epoch of accretion. This also serves to constrain the final rotation of the envelope to potentially large, but finite values.

For these studies to have any relevance to Betelgeuse, it is important that the structure remain that of an RSG after the proposed merger. As mentioned in \S \ref{subsec:cee}, a ``quiet merger" can leave behind an RSG, depending on pre-merger conditions. \citet{IvanovaPod03} suggest that this condition favors secondary masses $>2$ \msun\ and a primary close to carbon ignition so that strong gradients inhibit core/envelope mixing. \citet{Ivanova16} note that during a slow spiral-in, the angular velocity becomes constant in most of the CE and the value of the angular velocity is significantly smaller than the local Keplerian velocity in the envelope, so the approximation of spherical symmetry is reasonable.

\subsubsection{Angular Momentum}
\label{subsec:angmom}

During the merger and redistribution of density, angular momentum, and composition, some angular momentum is lost to the surroundings in the rotation-enhanced wind, and some is retained to propagate inward toward the primary core. In the 1D ``accretion" models explored by \cite{Sull20}, the angular momentum that is retained is redistributed by an inward diffusive wave of angular momentum. The profiles of the specific angular momentum and angular velocity quickly evolve to stable forms delineated by an inward propagating front with the specific angular momentum increasing outward beyond the front and the angular velocity being nearly constant.

A few years after accretion, the ingoing wave of angular momentum propagated to the boundary between the outer envelope and the H/He shell. The wave of angular momentum was halted at the composition boundary at the edge of the helium core leaving behind an envelope of constant angular velocity and a monotonically rising angular momentum per unit mass. By the epoch of collapse, the angular momentum distribution in the outer envelope had scarcely changed. The wave of angular momentum swept through the H/He shell, but was halted at the outer boundary of the He shell at 7 \msun\ for both the 20\msun+1\msun\ and the 20\msun+10\msun\ models. The composition distribution remained virtually unchanged.

All the final models of \cite{Sull20} have inner regions of negative gradient in $j$ in regions of sharp composition gradients. These must be stabilized against the Rayleigh instability by the associated composition gradients. This condition has not been investigated in detail.

\citet{Ivanova16} presented a model of a primary of 1.8 \msun\ and secondary of 0.1 \msun (model M10; their figure 7). While the mass scale is smaller than considered by \cite{Sull20} and \cite{chatz20}, the mass ratio, $\sim 0.05$, is about the same as for their 20\msun+1\msun\ models. The angular velocity as a function of mass for model M10 50 days after the plunge-in is basically flat throughout the model. The value of the angular velocity, $\sim 3\times 10^{-7}$ rad s$^{-1}$, is close to that of \cite{Sull20}, perhaps fortuitously, but the peak value of the angular momentum per unit mass for model M10 is about a factor of 30 less than found by \cite{Sull20}. The flat angular velocity profile in the 3D simulations seems to arise naturally in \textsc{mesa} simulations.

Significant departures in behavior between \citet{Ivanova16} and \cite{Sull20} are found in the innermost and the outermost regions. \citet{Ivanova16} do not consider the inner core, so they do not explore the distribution of angular momentum in the core. On the other hand, \citet{Ivanova16} and the upper right panel of Figure \ref{fig:postmerger} find a distinct decrease in both the specific angular momentum and the angular velocity in the outer 10 \% of the mass of the envelope that the models of \cite{Sull20} do not reveal. This difference probably arises in the loss of mass and angular momentum in the dynamical plunge-in phase that \cite{Sull20} do not treat accurately.

\subsubsection{Composition}
\label{subsec:comp}

In their 1D calculation with a fully-resolved primary core, \cite{Sull20} found the composition distribution of the inner core to be only slightly affected even by the accretion of a companion of large mass. Thus while the inner structure might be somewhat perturbed by accretion of substantial mass, there may be rather little effect on the outside to indicate that the accretion occurred. The implication is that the inner composition structure of Betelgeuse could be rather different depending on the mass accreted with basically no indication reflected in the outer, directly observable structure.

Evidence of internal mixing due to a merger (or other effects) can be revealed by anomalous surface abundances (\S \ref{subsec:abundances}). \citet{chatz20} were able to reproduce surface abundances of Betelgeuse, especially the observed overabundance of nitrogen \citep{lambert84}. \citet{brott11a, brott11b} and \citet{ekstrom12} find N/C surface enhancements that are similar to those found in the merger simulations of \citet{chatz20} due to the enchanced rotation from the spiraling-in phase.

\subsubsection{Entropy}
\label{subsec:entropy}

\citet{Ivanova16} give an extensive discussion of the treatment of entropy in CEE. They argue that the evolution of the entropy of the common envelope material differs between 3D and 1D simulations. In 1D, the entropy is generated because the energy is added as heat.  Since the radius at which the recombination energy release overcomes the potential well depends on the entropy of the material, the entropy generation observed in 1D codes will likely predict different outcomes than 3D CE evolution. \citet{Ivanova16} argue that 1D stellar codes should add the energy as mechanical energy rather than ``heat" that moves the material to a higher adiabat.

\citet{chatz20} find relatively little heating effects in their 3D merger simulation of Betelgeuse. We note, however, that heating during merger can lead to non-linear envelope pulsations and to potentially large mass loss \citep{Clayton17}.

\subsubsection{Recombination}
\label{subsec:recomb}

The role of hydrogen and helium recombination in abetting CE mass loss is discussed by \citet{IvanovaJP}, \citet{Ivanova16}, and \citet{lau22}. These reservoirs of energy can help to trigger envelope instability depending on where and when the recombination energy is released. The time-scale of recombination runaway can be up to several hundred days and gets longer as the mass of the companion decreases. In such cases, radiative losses can become important so that 3D simulations that lack radiative transfer are no longer appropriate. For all their limitations, 1D stellar evolution codes like \textsc{mesa} can handle this aspect of the physics.

\subsubsection{Magnetic Fields}
\label{subsec:mag}

As noted in \S\ref{subsec:magnet}, Betelgeuse displays various effects of magnetic fields. The magnetic properties are often omitted in CEE simulations. \cite{Sull20} and \citet{chatz20} included magnetic effects as treated by the \textsc{mesa} Spruit/Tayler algorithm in some cases, but did not include magnetic effects of the magnetorotational instability \citep{WKC15,moyanoMRI}. The omission of the latter will undoubtedly alter the quantitative, if not qualitative results. The Spruit/Tayler mechanism gives results that typically weight the radial component, $B_r$, orders of magnitude less than the toroidal component, $B_\phi$. The magnetorotational instability tends to give the radial component about 20 per cent of the toroidal component. Another important caveat is that \textsc{mesa} computes the magnetic field structure based on the instantaneous structure of the model. In reality, the field only decays on a dissipation timescale that might in some circumstances be long compared to the evolutionary timescales. This would lead to fossil magnetic field in a region that made a transition from being unstable to stable to the Spruit/Tayler instability. \textsc{mesa} has no means to treat the existence and decay of such fossil fields. The magnetic structure computed by \cite{Sull20} is thus interesting, but should not be given any quantitative weight.

\cite{Sull20} found that accretion has little effect on the production of magnetic fields by the Spruit/Tayler mechanism. Their models show a more substantial field in the outer part of the helium shell, reaching up to the base of the hydrogen envelope. The peak fields are of order 1 G and 1000 G for the radial and toroidal fields, respectively, with considerable variation with radius that is likely to be affected by issues of numerical resolution. Below an inward gap where the fields are very small the fields become large, but variable, in the innermost layers of the oxygen core. The radial fields peak at $\sim$ 1000 G and the toroidal fields at $\sim 10^6$ to $10^7$ G. In the models, the fields peak off center and the toroidal field declines to about 1 G in the center. The accretion appears to have a quantitative, but not qualitative, effect on the field strength and distribution just prior to collapse. Subsequent core collapse by a factor of $\sim 100$ in radius would amplify the field by compression alone by a factor of $\sim 10^4$. The resulting field of $\sim 10^{11}$ G would not be dynamically significant, but would give ample seed field for growth of the field in the proto-neutron star by the MRI \citep{Akiyama03, Obergaulinger09, Moesta18}

\subsection{Lessons from SN~1987A}
\label{subsec:87a}

To account for the circumstellar nebular rings, many studies of the mergers of massive stars have focused on the prospect that the progenitor of SN~1987A may have undergone a merger \citep{MorrisPod07}. Merger models can also account for why the progenitor was a blue rather than red supergiant by invoking mixing of helium from the core into the outer envelope \citep{MenonHeger17}.

In the case of Betelgeuse, a contrasting conclusion applies. While some discuss the possibility that Betelgeuse will explode as a blue supergiant \citep{vanloon2013}, Betelgeuse is still a red supergiant. If one accepts the basic {\it ansatz} that a merger is required to account for the observed rotational velocity of Betelgeuse, then it follows that a merger did not produce a compact blue envelope and thus, by the arguments of \citet{IPS02} and \citet{MenonHeger17}, little to no helium could have been mixed outward from the core, consistent with the simulations of \cite{Sull20} and \cite{chatz20}.

The modeling of a putative Betelgeuse merger by \citet{chatz20} (\S \ref{subsec:merger}) concluded that the plume from the disrupted secondary would not penetrate the helium core and induce substantial helium mixing according to the prescription of \citet{IPS02}. Mixing may be more likely for more massive secondaries, so the results of \cite{Sull20} and \cite{chatz20} may be less reliable for larger mass secondaries. Plume mixing is a complex hydrodynamical problem that deserves more study if we are to understand both Betelgeuse and SN~1987A as products of massive star mergers.

\section{The Great Dimming}
\label{sec:dimming}

While Betelgeuse is known to display a wide range of fascinating behavior, it surprised the Betelgeuse community and fascinated people world wide when it went through a phase of anomalously low optical luminosity beginning in October 2019 and lasting through March 2020 that became known as the Great Dimming. The brightness decreased by over a factor of two and was easily noticed by even casual observers of the night sky. Twitter alit with rampant speculation that Betelguese was about to explode, thus requiring some effort by supernova experts to tamp that particular hyped fever.

\begin{figure*}[htb!]
    \centering
    \includegraphics[width=0.99\linewidth]{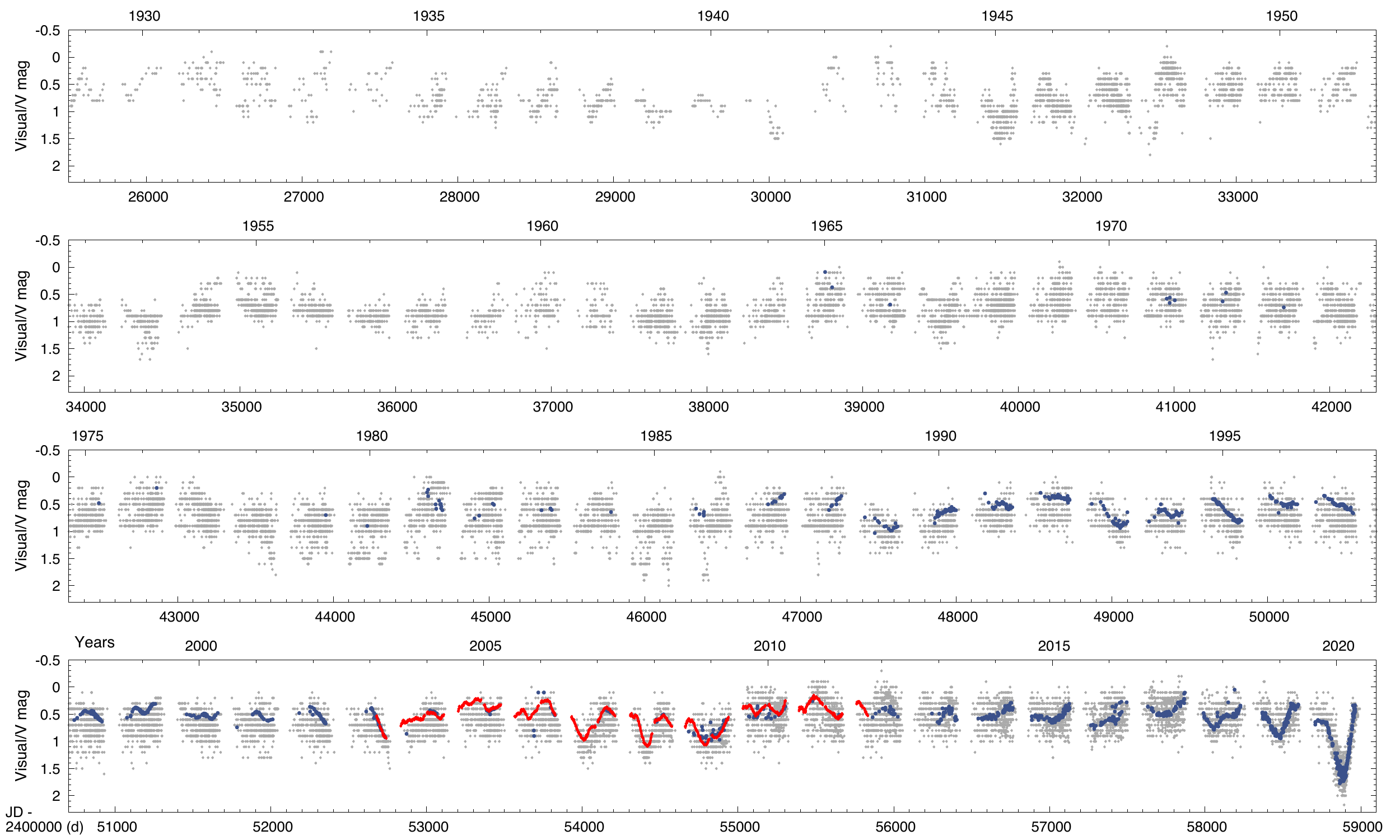}
    \caption{Century long record of the visual and V band brightness of Betelgeuse compliled by the American Association of Variable Star Observers supplemented by data from the Solar Magnetic Ejection Imager (SMEI). The final large dip is the Great Dimming of 2020 (\S \ref{sec:dimming}). From \citet{Joyce20} by permission of Meridith Joyce, L{\'a}szl{\'o} Moln{\'a}r, and the Astrophysical Journal.
}
    \label{fig:pulse}
\end{figure*}

The Great Dimming was dramatic enough on its own \citep{Guinan20,Levesque20,Harper20a,Dupree2020,dharma20,Harper20b,safonov20pol,montarges21,Levesque21,2021AJ....162..246H,2021NatCo..12.4719A,2021A&A...650L..17K,Dupree22,2022ApJ...934..131M,2022NatAs...6..930T,cannon23}. Both professionals and amateurs had monitored the brightness of Betelgeuse for about a century. Much of that data is stored in the valuable records of the American Association of Variable Star Observers (AAVSO) (Figure \ref{fig:pulse}). These studies had established that Betelgeuse was a variable star with regular pulsations on a variety of time scales as discussed in \S \ref{subsec:pulsation}. The decrease in V band amplitude corresponding to the $\sim 400$ day pulsation is typically 0.3–0.5 mag. In the Great Dimming, the decrease was over 1 mag.

There is some suggestion in the AAVSO records that Betelgeuse underwent other periods of anomalously large dimming, perhaps in the early 1950s and the late 1980s \citep{Joyce20}. Dimming with a period of about 30 - 40 years might be related to the rotation period of Betelgeuse (\S \ref{subsec:rotation}), but uncertainties in the early photometry and in the radius and rotational velocity and hence the period of Betelgeuse make that difficult to determine. There is also some concern that individual AAVSO observers in the 1950s and 1980s showed a tendency to report anomalously faint data that might have biased the mean AAVSO values and produced a false impression of minima (T. Calderwood, private communication, 2023).

There was no missing the Great Dimming of 2019/2020, but the attention to it may have been amplified by an initiative of Andrea Dupree of the Center for Astrophysics who convened many of the world's experts on Betelgeuse to participate in an intense global, multi-instrument, multi-wavelength coordinated study of Betelgeuse beginning in April, 2018, when Betelgeuse was especially well situated for studies with the HST, a project she named the Month of Betelgeuse (MOB). Forty-four astronomers joined the MOB. A month was not, of course, sufficient to address all the mysteries of Betelgeuse, the project was rapidly renamed Months of Betelgeuse. Because of the MOB activity, attention to Betelgeuse was still focused as the Great Dimming got underway. \citet{Dupree2020} witnessed a UV enhancement from Sept-Nov. 2019 with HST that may have been a precursor event to the Great Dimming prior to any substantial decrease in brightness.

Ed Guinan of Villanova had been doing careful photometry and other studies of Betelgeuse for over 25 years. \citet{Guinan19} reported V-band, Wing TiO-band, and NIR photometry on 7 December, 2019. They
presented one of the first interpretations of the Great Dimming, noting that “The light variations are complicated and arise from pulsations as well from the waxing and waning of large super-granules on the star's convective surface.”
They predicted a minimum on 21 February, 2020 $\pm$ a week, based on the assumption that a 420 day period was in a fortuitous concatenation with longer-term (5 - 6 yr) and shorter term (100 - 180 d) brightness changes and perhaps a super-granule upwelling of a cool plume. This estimate turned out to be remarkably close to the observed light minimum of $1.614\pm0.008$ mag during 07-13 February 2020 \citep{Guinan20}.

\citet{Levesque20} reported optical spectroscopy on 15 February, 2020. They also examined the TiO bands and reported a $T_{eff}$ of 3600 K compared to a typical value $\sim$ 3660 K. They argued that the small change in $T_{eff}$ was not commensurate with the large change in V-band luminosity and that a temporary cool period on the surface of Betelgeuse due to convective turnover was likely not the primary cause of the Great Dimming. Rather, they proposed an increase in large-grain gray dust.

\citet{Guinan20} and \citet{Levesque20} thus framed the extremes of the possible explanations of the Great Dimming that continues to be debated. Early discussions of $T_{eff}$ during the Great Dimming assumed spherical symmetry. While the assumption of spherical symmetry was an obvious starting point, subsequent consideration of spots, super-granules, and circumstellar dust call that assumption into question. The question of how accurately a global value of $T_{eff}$ can be determined if $T_{eff}$ varies over the surface of the star remains to be determined.

\begin{figure*}[htb!]
    \centering
    \includegraphics[width=0.99\linewidth]{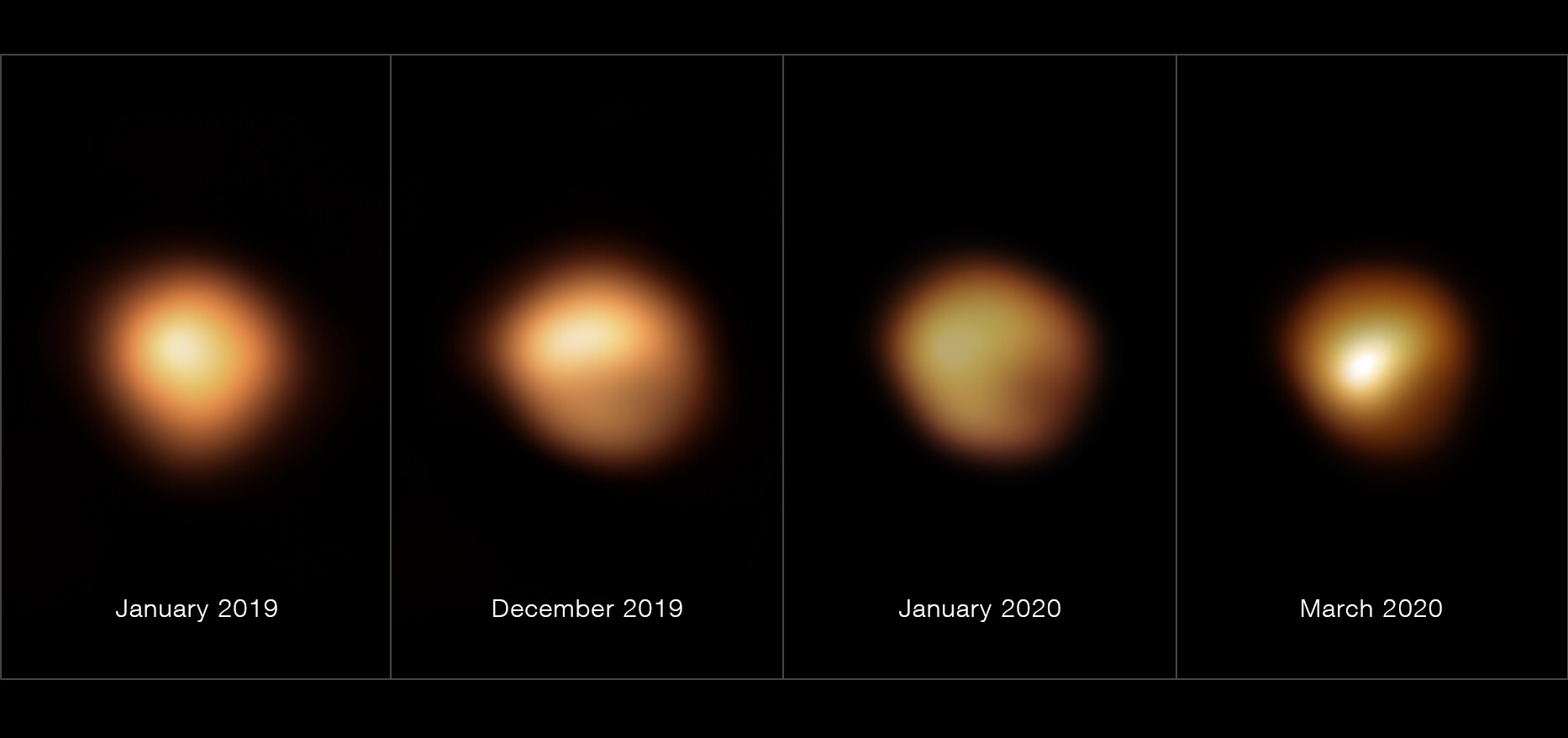}
    \caption{Resolved images of Betelgeuse through the Great Dimming. January 2019 was prior to the dimming; December 2019 was somewhat after the onset of the dimming; January 2020 was near the minimum of the dimming; and March 2020 was after the maximum dimming. From \citet{montarges21} by permission of M. Montarg{\`e}s and ESO.}
    \label{fig:montarge}
\end{figure*}

\citet{montarges21} illustrated this conundrum by producing dramatic spatially-resolved interferometric images of Betelgeuse obtained with the SPHERE instrument on the VLT in December 2019 and January 2020 that showed that the southern half of the star had become markedly fainter than in January 2019, indicating that a major change has occurred in, or near, the photosphere (Figure \ref{fig:montarge}). \citet{montarges21} attributed this dark patch to ``a dusty veil."

\begin{figure}[htb!]
    \centering
    \includegraphics[width=0.99\linewidth]{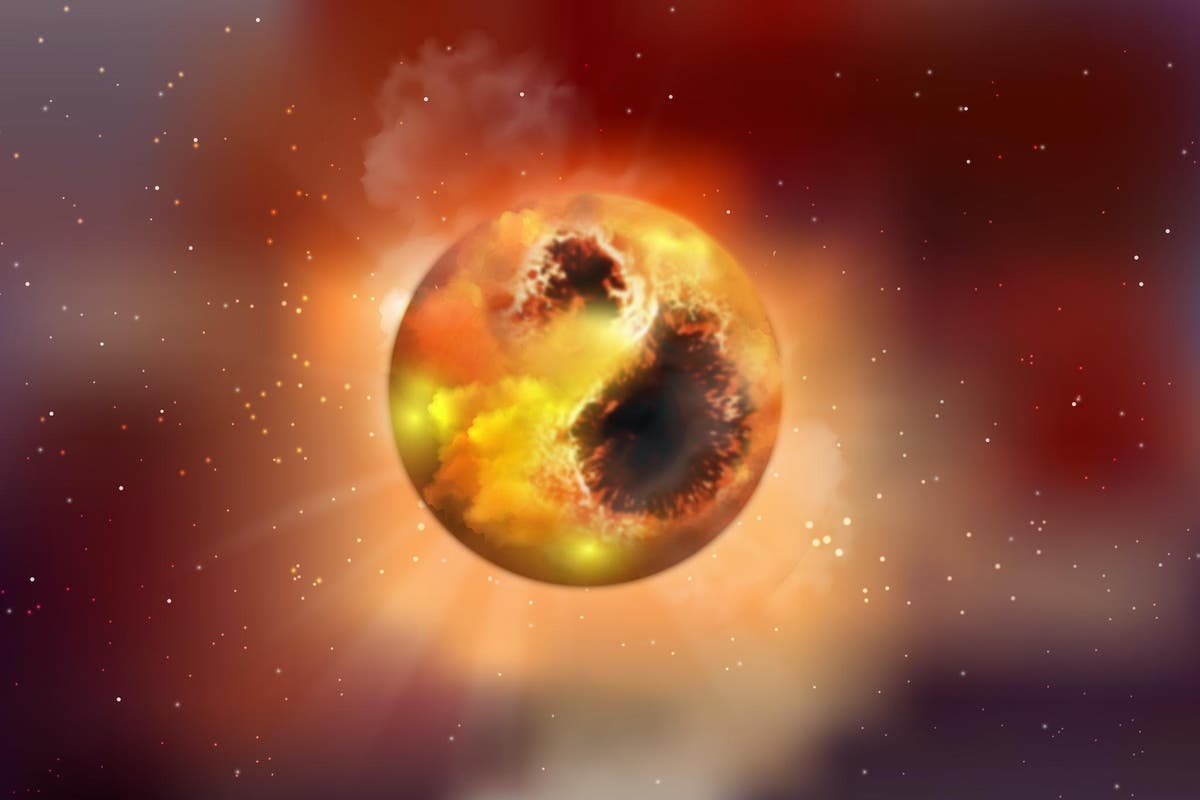}
    \caption{Schematic of gigantic cool dim patches on the surface of Betelgeuse that were proposed as contributing to the Great Dimming. By permission of T. Dharmawardena and Max Planck Institute for Astrophysics, Heidelberg. Credit: Judith Neidel, MPIA graphics department.
}
    \label{fig:patches}
\end{figure}

\citet{Harper20b} reported Wing three-filter (A, B, and C band) TiO and near-IR photometry that showed that portions of the photosphere had a mean $T_{eff}$ that was significantly lower than that found by \citet{Levesque20}. They interpreted the image of \citet{montarges21} to be a large patch in the photosphere that could be 250 K cooler than surroundings. They concluded that no new dust was required and emphasized the interpretation of \citet{Guinan19} that the Great Dimming resulted from a coincidence of the 430 day and 5.8 year periods of Betelgeuse. They suggested that the cooling of a large portion of the surface was produced dynamically by photospheric motions due to pulsation and large-scale convective motions (Figure \ref{fig:patches}).

\citet{dharma20} reported 13 years of submillimeter observations of Betelgeuse including the epoch of the Great Dimming obtained with the James Clerk Maxwell Telescope and the Atacama Pathfinder Experiment. These long wavelength observations were significant because they were not expected to be obscured by dust as the optical observations could be.  \citet{dharma20} found that Betelgeuse had also dimmed by $\sim$ 20\% at these longer wavelengths during the optical minimum. They concluded that the dimming was due to changes in the photospheric luminosity as opposed to obscuration by surrounding dust. See also \citet{2022ApJ...934..131M} for 1.3 cm and 7 mm observations with the VLA on August 2, 2019, just prior to the onset of the optical dimming and \citet{2021AJ....162..246H} for observations of circumstellar [O I] 63.2 $\mu$m and [C II] 157.7 $\mu$m emission profiles and [O I] 63.2 $\mu$m, [O I] 145.5 $\mu$m, and [C II] 157.7 $\mu$m fluxes obtained shortly after the Great Dimming with SOFIA.

\begin{figure}[htb!]
    \centering
    \includegraphics[width=0.99\linewidth]{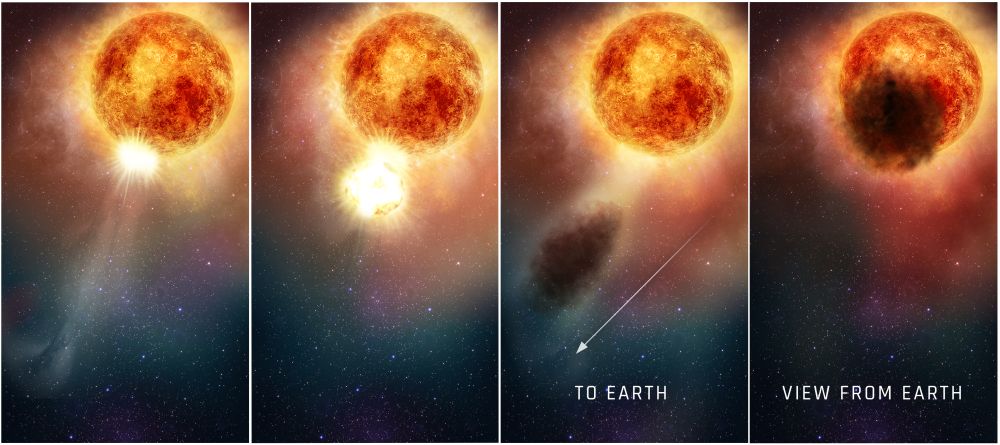}
    \caption{Schematic of model of the Great Dimming in which a blob of gas is ejected, cools to form dust, and passes through the line of sight to cause the Great Dimming. Courtesy Ray Villard, STScI and NASA, art by E. Wheatley, STScI.
}
    \label{fig:blob}
\end{figure}

\citet{Dupree22} presented a synthesis of observations and an interpretation of the Great Dimming that was a variation on the picture first presented by \citet{Kervella18}. \citet{Dupree22} outlined a scenario in which the UV burst reported by \citet{Dupree2020} represented a surface activity perhaps catalyzed by an upwelling that ejected a blob of matter. That matter cooled as it expanded away from the surface allowing dust to form. The resulting clump of dust crossed the line of sight to Betelgeuse resulting in the dim patch observed by \citet{montarges21} and the resulting Great Dimming (Figure \ref{fig:blob}).

\citet{2021NatCo..12.4719A} presented high-resolution high S/N ratio near-infrared spectra obtained at Weihai Observatory on four epochs in 2020 during and after the Great Dimming. They argued that a decrease in the overall mean $T_{eff}$ by at least 170 K on 2020 January 31 could be attributed to the emergence of a large dark spot on the surface of Betelgeuse. \citet{Levesque21} argued for a synthesis in which the dark patch of \citet{montarges21} and the Great Dimming were caused by dust forming over a cold patch in the southern hemisphere. \citet{cannon23} presented VLTI  Multi AperTure mid-Infrared SpectroScopic Experiment (MATISSE) observations in the N-band (8 - 13 $\mu$m) near brightness minimum. They explored a model invoking multiple clumps of dust close to the star and another considering a large cool spot on the stellar surface with no dust. They found that both the dust clump and the cool spot models are compatible with the data, noting that the extinction and emission of a localised dust clump in the line of sight compensate each other making the clump undetectable. They concluded that the lack of infrared brightening during the Great Dimming \citep{dharma20} does not exclude extinction due to a dust clump as one of the possible mechanisms.

\begin{figure*}[htb!]
    \centering
    \includegraphics[width=3.5in]{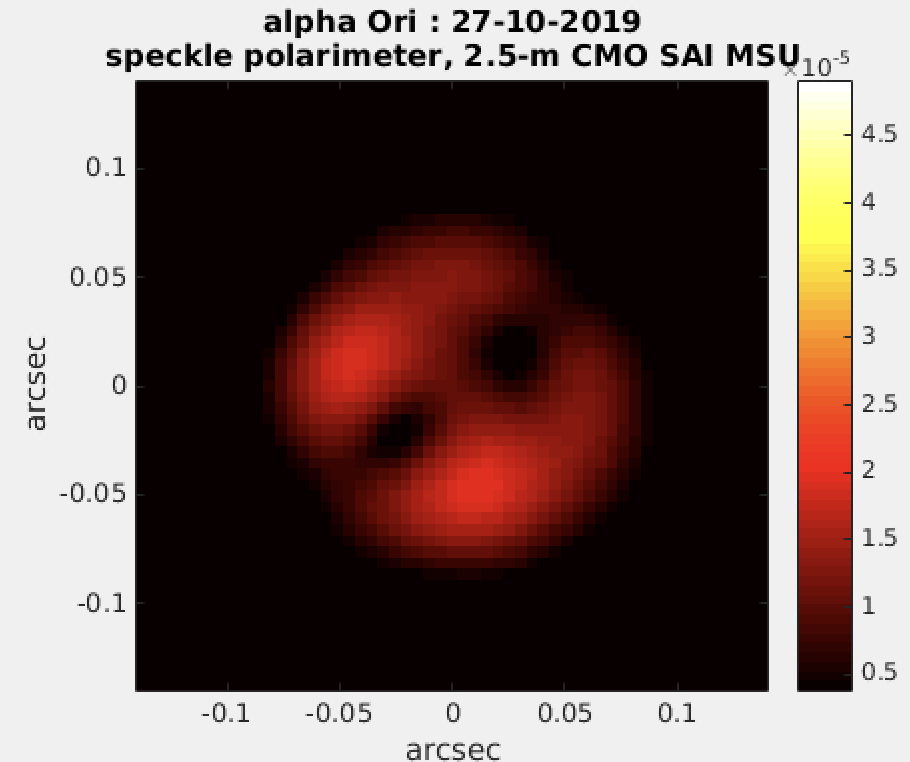}
    \includegraphics[width=3.5in]{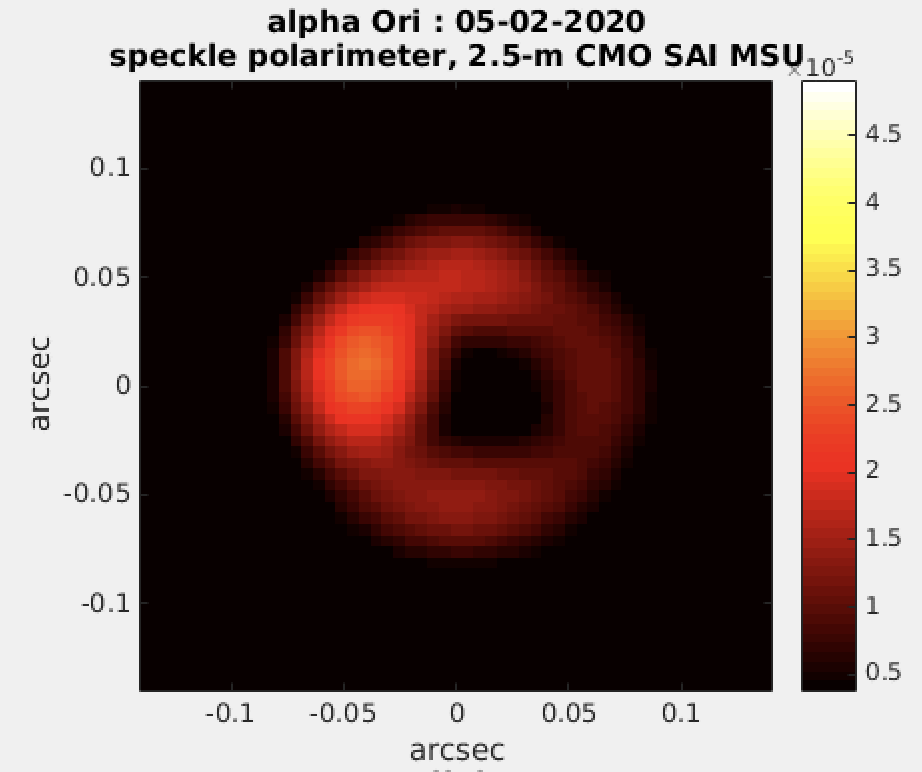}
    \caption{Speckle polarimetry of the surface of Betelgeuse on 27 October 2019 at the very beginning of the Great Dimming (left panel) and on 5 February 2020 at the extreme of the dimming (right panel). From \citet{safonov20pol} by permission of B. Safanov, Caucasian Mountain Observatory, and the Sternberg Astronomical Institute of Moscow State University.
}
    \label{fig:pol}
\end{figure*}

Spectropolarimetry provides yet another tool to explore the geometry of the surface of Betelgeuse \citep{lopezpol,haubois19,2020RNAAS...4...39C}. \citet{safonov20pol} argued that to address the challenges of the Great Dimming it was fundamentally important to employ methods to resolve an inhomogeneous stellar atmosphere. They presented a set of differential speckle polarimetric observations of Betelgeuse obtained at the 2.5 m telescope of the Caucasian Mountain Observatory operated by the Sternberg Astronomical Institute of  Moscow State University. Observations on 17 days at wavelengths 465, 550, 625 and 880 nm spanned the Great Dimming event. The envelope was found to be highly inhomogeneous but correlated, with features varying on a timescale of two or three months (Figure \ref{fig:pol}). An animation captured the dramatic variablility registered in the polarization \footnote{\url{http://lnfm1.sai.msu.ru/kgo/mfc\_Betelgeuse\_en.php}}.
The net polarized brightness of the envelope remained constant as the $V$ band flux approached its minimum. After the minimum, the polarized flux of the envelope rose a factor of two as the optical flux was restored. \citet{safonov20pol} concluded that the Great Dimming was caused by the formation of a dust cloud located on the line of sight.


\citet{2021A&A...650L..17K} used high resolution spectroscopy to do a tomographic study of the structure during the Great Dimming. This analysis suggested that two shocks propagated in the upper atmosphere, one generated in February 2018 and one in January 2019 with the second amplifying the effects of the first. \citet{2021A&A...650L..17K} suggest that this shock structure modified by underlying convection or outward gas motion altered the molecular opacity in the line of sight.

\citet{mittag23} have engaged in long-term studies of the chromosphere of Betelgeuse since 2013, including the epoch of the Great Dimming when they determined the absolute and normalized excess flux of the Ca II H\&K lines. They found a behavior similar to that during a previous decrease in the visual brightness of Betelgeuse in 1984 and 1985. Unlike the Mg II emission \citep{Dupree22}, the Ca II emission attributed to the lower chromosphere of Betelgeuse did not change significantly between November 2019 and February 2020, but did vary after the Great Dimming. \citet{mittag23} argue that this delay of the chromospheric reaction suggests that the cause for the great dimming is located in the photosphere.

Himawari-8 is a Japanese geostationary meteorological satellite orbiting 35,786~km above the equator at 140.74\degree E. Several bright stars, including Betelgeuse, occasionally fortuitiously appear in the images. \citet{2022NatAs...6..930T} used the 16-band photometry from 0.45 to 13.5 $\mu$m from Himawari-8 to construct a light curve of Betelgeuse spanning 4.5 yr from 2017 to 2021 with a sampling that averaged once per 1.72 d. \citet{2022NatAs...6..930T} used the infrared optical depth in contrast to variations in the visual extinction, A(V), to directly trace the amount of circumstellar dust. The IR optical depth increased during the Great Dimming with a small delay between the peak in IR optical depth and the extinction in the visual. \citet{2022NatAs...6..930T} argue that their data suggest that a clump of gas produced dust that obscured the photosphere of Betelgeuse and contributed to the Great Dimming and that the enhancements of visual extinction and IR optical depth during the Great Dimming may have occurred very close to the photosphere. They conclude that their results support a scenario in which the Great Dimming was caused by a combination of a decrease in $T_{eff}$ and an increase in A(V) in roughly equal amounts consistent with the change in polarization reported by \citet{2020RNAAS...4...39C} and \citet{safonov20pol}.

The Great Dimming was clearly a complex phenomenon. It seems unlikely that its alignment with the phase of the $\sim 400$ d pulsation period was a coincidence, but that alone cannot account for the magnitude and spatial inhomogeneity of the obscuration. There seems to be solid evidence that dust with an inhomogeneous spatial distribution played a role. A decrease in $T_{eff}$ occurred, perhaps in the form of large, cooler patches. Such patchy structure in surface temperature calls into question the meaning of a global $T_{eff}$ and at least sets upper limits on the precision with which a global $T_{eff}$ can be determined.

In an epilogue, after the Great Dimming, Betelgeuse has shown repeated minima at an interval of $\sim 200$ d and has steadily increased in mean brightness to reach an historic maximum of V $\sim 0$ in April, 2023 (M. Montarg{\`e}s, private communication, 2023).

\section{The Explosion to Come}
\label{sec:explo}

Betelgeuse will eventually explode, most probably still as a red supergiant and thus producing a Type II supernova \citep{snex}. The details will be affected by the intense convection in the late shell-burning phases less than a year before explosion that may also produce significant outward acoustic flux and mass loss. The convection seeds turbulence in the collapsing material that is expected to enhance the subsequent explosion \citep{arnett11,couch15,chatz16}.

At the expected iron-core collapse, $10^{53}$ ergs of neutrinos will be produced that emerge from the envelope about an hour after collapse and flood into space. About 600 years later, detection of those neutrinos will give humans on Earth (if there are any 100,000 + 600 years from now) their first hint of the events to come. At that time, a human body would receive $\sim$ 100 trillion neutrinos, vastly less than a lethal dose of radiation.

The shock wave generated by the explosion carrying $\sim10^{51}$ erg of kinetic energy will take about a day to reach the surface. A blast of UV lasting about an hour will then occur. The resulting UV flux will be less than the flux of the Sun at Earth but perhaps sufficient to cause some disruption of atmospheric chemistry.

In two weeks, the explosion will be producing a billion times the solar luminosity. At the Earth, Betelgeuse will appear as a pinpoint about as bright as a quarter Moon lasting for $\sim$3 months. The explosion will then fade, but remain visible at the Earth for many years and to scientific instruments for centuries. The explosion of Betelgeuse is likely to produce a pulsar that will also be visible for a million years or more.

The supernova blast wave will propagate out through the surrounding CSM and ISM and interact with any mass lost in the next 100,000 years and eventually with the complex CSM illustrated in Figure \ref{fig:csm}. The shock wave will propagate at $\sim 5000$~\kms\ and hence collide with the bow shock in about 60 y and the odd linear structure about 20 y later, assuming a distance to Betelgeuse of 165 pc \citep{Joyce20}.

By the time the supernova shock reaches the Earth more than 100,000 years after the explosion, the Solar magnetosphere should easily deflect it. The Earth immersed within the supernova remnant may witness an increase in cosmic ray flux.

\section{Summary, Conclusions, and Future}
\label{sec:conclusion}

As nearby and well studied as it is, Betelguese still presents a host of outstanding issues: its irregular surface; the manner in which it ejects matter to form a chromosphere, wind, dust, and molecules; how it came to move through space and spin so rapidly; the nature of its variability and magnetic fields; and the possibility that it underwent a significant change in color within historical times. Statistics suggest that it was likely to have been born in a binary system and undergone complex common envelope evolution. Are the distant circumstellar structures related to the interaction of winds with the interstellar medium or the products of the dramatic turmoil of a merger?

A pertinent question is the structure and condition of Betelgeuse as we see it today, gracing Orion. While uncertainties in the distance remain troubling, Betelgeuse is most likely near the tip of the RSB. Since core helium burning is far longer than subsequent burning phases, Betelgeuse is most likely in core helium burning. The pulsation period likely constrains the radius and distance and the evolutionary state to core helium burning \citep{Joyce20}, but there are arguments to the contrary \citep{Neu22,lau22}.

Estimates of the surface gravity of Betelgeuse span a rather large range. The inclination angle is uncertain. A more accurate measurement of $log~g$, radius, and distance could yield new constraints on the current total mass. Comparison to the luminosity that most directly measures the core mass could reveal hints of the current mass of the outer hydrogen envelope, past mass loss, and the current evolutionary state. Continued exploration of atomic, isotopic, and molecular abundances may yield more information on internal mixing processes.

The notion that Betelgeuse may have undergone a merger remains viable. The extra angular momentum may have come from merger with a companion in the red supergiant phase, nearly independent of the mass of the secondary. Once a transient phase of merging has settled down and substantial mass and angular momentum have been ejected, there is rather little external difference in models in late core helium burning and subsequent phases. This frustrates attempts to determine the internal structure and state of the star. How can we prove Betelgeuse has undergone a merger?

There remain great challenges in understanding the associated physical processes if Betelgeuse underwent a merger. These must be explored with extensive, highly--resolved 3D studies of the formation and evolution of tidal plumes as any secondary is disrupted near the core of the primary. How deeply does the plume penetrate? What abundances are mixed to the surface? What is the level of envelope enrichment with helium that may determine whether the star remains red or moves back to the blue? What is the expected structure of the inner core as it evolves, rapidly rotating, beyond core helium burning?  

The Great Dimming of 2019/2020 did not portend imminent explosion. The origin of the dimming might have been related to a resonance of pulsation periods, expulsion of a dust cloud, and extra large star spots.

How can we test the current evolutionary state of Betelgeuse with observations of pulsations and surface irregularities? Are there faint signals from acoustic waves generated internally that carry information about the core structure that would prove or disprove the hypothesis of a merger?

What are the clues to imminent collapse that might pertain now, or in the far future if Betelgeuse is in core helium burning now? Models predict extensive mass loss shortly before collapse driven by rapid convection in the inner burning shells and there are hints of such pre-collapse mass loss in Type II supernovae that are thought to arise in red supergiants like Betelgeuse.

Mysteries abound!

\section{Addendum: Post-publication Notes}

Andrea Dupree’s Month of Betelgeuse project, MOB, that helped to focus attention on the Great Dimming was promptly renamed Months of Betelgeuse as the work of the group continued after April 2018.

Tom Calderwood (private communication, 2023) pointed out that possible deep minima in the light curve of Betelgeuse registered in AAVSO visual estimates in the early 1950s and mid to late 1980s came from only one or two amateur observers. Those data and hence the depth of any minima may be suspect.

After the Great Dimming, Betelgeuse brightened to near historic highs, oscillating with a dominant timescale closer to 200 days rather than 400 days (Sigismondi et al. 2023; Astronomer Telegram \#16001: Monitoring Betelgeuse at its brightest).

Saio et al. (2023; arXiv:2306.00287) presented an argument that the 2000 day “long secondary period” of Betelgeuse was the fundamental radial oscillation period, and that Betelgeuse may already be in carbon burning with only years to live. Both these points of view were contested by Molnár et al. (2023; Research Notes of the AAS, Volume 7, Number 6).


\section*{Acknowledgments}

We are grateful to Natasha Ivanova for discussions of common envelope evolution and to the Aspen Center for Physics for providing the environment to do so. We also thank Ed Guinan, Meridith Joyce, and Andrea Dupree and the Month of Betelgeuse (MOB) team for discussions of Betelgeuse and mergers. We thank Ralph Neuhauser, Brad Schaefer, and Anita Richards for discussions of the historical color evolution of Betelgeuse. We are especially thankful for the ample support of Bill Paxton and the \textsc{mesa} team. JCW is grateful to the group of enthusiastic undergraduates at The University of Texas at Austin who catalyzed his interest in Betelgeuse: Sarafina Nance, Jamie Sullivan, Manuel Diaz, Steven Smith, Justin Hickey, Li Zhou, Maria Koutoulaki, and Julia Fowler. The research of JCW was supported in part by the Samuel T. and Fern Yanagisawa Regents Professorship in Astronomy and by NSF AST-1813825. EC is grateful to Dr. Juhan Frank, Dr. Sagiv Shiber, and Dr. Bradley Munson for insightful discussions and for their collaboration. The research of EC was supported in part by the National Science Foundation Grant AST-1907617 and in part by the Department of Energy Early Career Award DE-SC0021228.

\software{\textsc{mesa} \citep{Paxton11, Paxton13, Paxton15,Paxton18}};
{\it OctoTiger} Adaptive Mesh Refinement (AMR) hydrodynamics code developed by the LSU Center for Computation and Technology (CCT) \citep{marcello2021}.


\bibliographystyle{mnras}

\bibliography{betelgeuse}

\end{document}